\documentclass[10pt,aps,floats,floatfix,nofootinbib,prd,superscriptaddress,twocolumn]{revtex4-2}
\usepackage{graphicx,mathtools,ragged2e,subcaption}
\usepackage[dvipsnames]{xcolor}
\usepackage[linktocpage,colorlinks=true,citecolor=OliveGreen,linkcolor=Maroon,urlcolor=Maroon]{hyperref}

\begin{document}

\title{Resonances in binary extreme mass ratio inspirals}
\author{Jo\~ao S. Santos}
\affiliation{CENTRA, Departamento de F\'{\i}sica, Instituto Superior T\'ecnico -- IST, Universidade de Lisboa -- UL,
Avenida Rovisco Pais 1, 1049-001 Lisboa, Portugal}
\affiliation{Center of Gravity, Niels Bohr Institute, Blegdamsvej 17, 2100 Copenhagen, Denmark}
\author{Vitor Cardoso}
\affiliation{CENTRA, Departamento de F\'{\i}sica, Instituto Superior T\'ecnico -- IST, Universidade de Lisboa -- UL,
Avenida Rovisco Pais 1, 1049-001 Lisboa, Portugal}
\affiliation{Center of Gravity, Niels Bohr Institute, Blegdamsvej 17, 2100 Copenhagen, Denmark}
\author{Alexandru Lupsasca}
\affiliation{Department of Physics \& Astronomy, Vanderbilt University, Nashville TN 37212, USA}
\affiliation{OpenAI}
\author{Jos\'e Nat\'ario}
\affiliation{CAMGSD, Departamento de Matem\'{a}tica, Instituto Superior T\'ecnico -- IST, Universidade de Lisboa -- UL,
Avenida Rovisco Pais 1, 1049-001 Lisboa, Portugal}
\author{Maarten van de Meent}
\affiliation{Center of Gravity, Niels Bohr Institute, Blegdamsvej 17, 2100 Copenhagen, Denmark}
\affiliation{Max Planck Institute for Gravitational Physics (Albert Einstein Institute), Am M\"uhlenberg 1, Potsdam 14476, Germany}

\date{\today}

\begin{abstract}
Stellar-mass binaries evolving in the vicinity of supermassive black holes (SMBHs) may be common in the universe, either in active galactic nuclei or in other astrophysical environments.
Here, we study in detail the resonant excitation of SMBH modes driven by a nearby stellar-mass binary.
The resulting resonant energy fluxes vary with the orbital location and frequency of the binary, exhibiting a rich and complex structure.
In particular, we find that the total energy flux radiated to infinity is maximized at a gravitational-wave frequency that is close to, but not exactly equal to, the real part of the corresponding quasinormal-mode frequency.
Moreover, as the binary is moved farther away from the SMBH, this offset from the mode frequency becomes increasingly pronounced.
In addition, for suitable orientations, the binary can effectively ``feed'' the light ring of the SMBH, selectively exciting particular oscillation modes.
For rotating (Kerr) black holes, the mode spectrum is significantly more intricate; however, individual modes are also less strongly damped, leading to an enhanced---but more difficult to interpret---resonant response.
\end{abstract}

\maketitle

%%%%%%%%%%%%%%%%%%%%%%%%%%%%%%%%%%%%%%%%%%%%
\section{Introduction}
%%%%%%%%%%%%%%%%%%%%%%%%%%%%%%%%%%%%%%%%%%%%

Black holes (BHs), like any other physical system, possess characteristic modes of oscillation, known as quasinormal modes (QNMs)~\cite{Chandrasekhar:1975zza,Berti:2009kk}.
These modes can be understood as being associated with gravitational waves (GWs) undergoing quasi-trapped motion around the BH---a phenomenon that occurs close to, but not at, the event horizon~\cite{Cardoso:2008bp,Dolan:2010wr,Yang:2012he}.
In other words, the proper oscillation modes of BHs are localized near the light ring: they encode information about the near-horizon region, yet do not directly discriminate the presence of horizons themselves~\cite{Cardoso:2019rvt}.
QNMs play a central role in virtually all dynamical processes involving BHs and their relaxation.
The ongoing BH spectroscopy program seeks to understand how perturbed BHs relax toward a final quiescent state~\cite{Berti:2005ys,Dreyer:2003bv,Berti:2009kk,Barack:2018yly,Cardoso:2019rvt,Berti:2025hly}.
The breakthrough detection of GWs from merging BHs has made this an especially vibrant area of research~\cite{LIGOScientific:2016lio,Cotesta:2022pci,LIGOScientific:2025rid,Berti:2025hly}.

It has recently emerged that astrophysical ``tuning forks,'' capable of self-tuning to the QNMs of supermassive BHs, may exist and may even be common in the cosmos~\cite{Bartos:2016dgn,Stone:2016wzz,Peng:2024wqf,Sberna:2022qbn,Dittmann:2023sha,Yang:2024tje,LIGOScientific:2025brd,Phukon:2025cky,Li:2025fnf,Stegmann:2025zkb}.
Such systems arise in the form of so-called binary extreme-mass-ratio inspirals (b-EMRIs): a stellar-mass binary orbiting a supermassive BH~\cite{Chen:2018axp,Han:2018hby,Yin:2024nyz,Jiang:2024mdl}. In general, the intrinsic orbital frequency of the stellar-mass binary increases due to drag or GW emission, and it may eventually be tuned to the QNMs of the supermassive BH~\cite{Cardoso:2021vjq}.

Although modeling a system as complex as a b-EMRI is highly challenging, recent progress has been made using Dixon’s formalism~\cite{Dixon:1970zza,Dixon:1974xoz}, which describes the stellar-mass binary as an effective particle with internal structure.
Retaining terms up to quadrupole order captures the generation of GWs from the internal motion of the stellar-mass binary~\cite{Santos:2025ass}.
In the same work, the emission and propagation of GWs were treated from first principles using BH perturbation theory~\cite{Santos:2025ass}.

Here, we employ the formalism and tools developed in Ref.~\cite{Santos:2025ass} to study the resonant excitation of the QNMs of the central supermassive BH.
Specifically, we place a stellar-mass binary in the vicinity of a supermassive BH, keeping its center of mass (CoM) static.\footnote{Formally, this configuration is not a solution of the Einstein field equations, since some external energy–momentum is required to prevent the binary from falling.
Nevertheless, keeping the binary static leads to a monochromatic GW signal, making resonances significantly easier to identify.
The fully dynamical case is left for future work.}
We find that the modes of the supermassive BH can indeed be resonantly excited.
However, this pattern of excitation exhibits a nontrivial dependence on the orbital frequency of the stellar binary, its distance from the supermassive BH, the inclination of its orbital plane, and the spin of the supermassive BH.

This paper is organized as follows.
First, in Sec.~\ref{sec:Piano}, we review the connection between BH QNMs and the light ring, interpreting the ring as a ``black hole piano.''
Then, in Sec.~\ref{sec:Framework}, we briefly describe the framework developed in Ref.~\cite{Santos:2025ass} to analyze b-EMRIs and present the setup studied here.
Next, we present our results in Sec.~\ref{sec:Results}, and finally conclude in Sec.~\ref{sec:Discussion}.
Throughout this work, we use geometric units with $G=c=1$.

%%%%%%%%%%%%%%%%%%%%%%%%%%%%%%%%%%%%%%%%%%%%
\section{Black hole piano} \label{sec:Piano}
%%%%%%%%%%%%%%%%%%%%%%%%%%%%%%%%%%%%%%%%%%%%

QNMs are the natural ringing tones of BHs.
When a BH is disturbed, it rings at complex QNM frequencies: their real part sets the oscillation frequency (the ``note''), while the imaginary part controls the damping rate (how fast the note dies away).
The QNM spectrum $_{(s)}\omega_{\ell mn}^{\rm QNM}$ is labeled by the spin-weight $s$ of a perturbation, harmonic numbers $(\ell,m)$ with $-\ell\le m\le+\ell$ and $\ell\ge|s|$, and an overtone number $n\ge0$.
Higher overtones are exponentially more damped than the fundamental $n=0$ mode.

In the high-frequency (``eikonal'') limit, massless waves propagating on a BH background can be approximated by congruences of null geodesics that trace out the wavefronts: this is the geometric optics approximation. 

In this eikonal regime, it is well-known that the QNMs correspond to congruences of null geodesics that asymptote to bound photon orbits around the BH.
For a Schwarzschild BH, there is a unique radius $\tilde{r}=3M$ at which light can orbit (the so-called ``photon sphere'' or ``light ring''), and the eikonal spectrum is \cite{Goebel1972,Ferrari1984,Mashhoon1985,Cardoso:2008bp}
\begin{equation}
    \omega_{\ell mn}^{\rm QNM}\stackrel{\ell\to\infty}{=}\left(\ell+\frac{1}{2}\right)\tilde{\Omega}-i\left(n+\frac{1}{2}\right)\gamma_L+\mathcal{O}\!\left(\frac{1}{\ell}\right)\,, \label{eq:EikonalQNM}
\end{equation}
where $\tilde{\Omega}=\frac{1}{3\sqrt{3}M}$ denotes the angular velocity of a bound photon orbiting at $\tilde{r}=3M$, while $\gamma_L=\frac{1}{3\sqrt{3}M}$ is the Lyapunov exponent governing its orbital instability.
In the eikonal limit, the frequency diverges as the total angular momentum, $\omega\sim\ell\to\infty$. The effects of spin-weight are $\mathcal{O}(1/\ell)$, so $s$ drops out at this approximation order.

In this spherically symmetric case, there is no dependence on the azimuthal mode number $m$, but in general, the eikonal limit is defined by taking $\ell,m\to\infty$ while keeping the ratio $\mu\equiv m/\ell\in[-1,1]$ fixed.

A quantitative analysis of the eikonal spectrum of Kerr QNMs was undertaken in Refs.~\cite{Yang:2012he,Hadar:2022xag}, which found that Eq.~\eqref{eq:EikonalQNM} continues to hold for the rotating BH, but with $\tilde{\Omega}$ and $\gamma_L$ functions of $\mu=m/\ell$.
These functions admit a beautiful geometric interpretation that we now briefly review (see, e.g., Sec.~2.7 of Ref.~\cite{Detournay:2025xqd} for more details).

A Kerr BH is surrounded by a ``photon shell'' of bound photon orbits spanning multiple radii $\tilde{r}\in[\tilde{r}_+,\tilde{r}_-]$ \cite{Teo:2020sey,Johnson:2019ljv}.
The prograde and retrograde circular-equatorial photon orbits at $\tilde{r}_\pm$ delineate the boundaries of the photon shell.
In its interior, bound null geodesics oscillate in polar angle $\theta$ between turning points $\theta_+(\tilde{r})$ and $\pi-\theta_+(\tilde{r})$.
Thus, each orbital radius $\tilde{r}\in[\tilde{r}_+,\tilde{r}_-]$ has its own ``signed inclination'' $\pm\sin{\theta_+(\tilde{r})}\in[-1,1]$: this is a bijective map whose sign is set by the azimuthal angular momentum of the orbit (i.e., whether it is prograde or retrograde).

In the geometric optics approximation, QNMs with a given $\mu=m/\ell$ correspond to photon orbits with inclination $\sin{\theta_+}=\mu$.
This defines a bijection between $\mu$ and orbital radius $\tilde{r}$ within the photon shell.
At large but still finite $\ell$, $\mu\in[-1,1]$ is not a continuous parameter, but rather takes $2\ell+1$ values corresponding to a discrete set of orbital radii: the ``strings'' of the BH piano.

Finally, BH images display a narrow, bright ``photon ring'' produced by light that orbited in the photon shell before escaping to our telescopes \cite{Johnson:2019ljv}.
The angle $\varphi(\tilde{r})$ around the photon ring corresponds to light bound at different orbital radii $\tilde{r}$, with different inclinations $\sin{\theta_+}$, and hence to different QNMs.
Eikonal modes with a given $\mu$ correspond to the same angle $\varphi(\mu)$.
For large but not infinite $\ell$, $\mu$ is discrete and thus the photon ring is ``quantized'' into $2\ell+1$ angles: the ``keys'' of the BH piano.
The photon ring in BH images is thus a picture of the piano's keyboard.

In summary, a SMBH is like a piano, whose strings (QNMs) consist of bound photon orbits, and whose keys lie around its photon ring, which is a BH keyboard.
It is a fundamental question to understand how these notes are played.
In the remainder of the paper, we use a small binary as a tuning fork, whose internal orbital motion produces high-frequency GWs and plays the BH piano.

%%%%%%%%%%%%%%%%%%%%%%%%%%%%%%%%%%%%%%%%%%%%
\section{Framework} \label{sec:Framework}
%%%%%%%%%%%%%%%%%%%%%%%%%%%%%%%%%%%%%%%%%%%%

%%%%%%%%%%%%%%%%%%%%%%%%%%%%%%%%%%%%%%%%%%%%
\subsection{The b-EMRI setup} \label{sec:b_EMRI}
%%%%%%%%%%%%%%%%%%%%%%%%%%%%%%%%%%%%%%%%%%%%
%
\begin{figure}[t]
\centering
\includegraphics[width=.45 \textwidth]{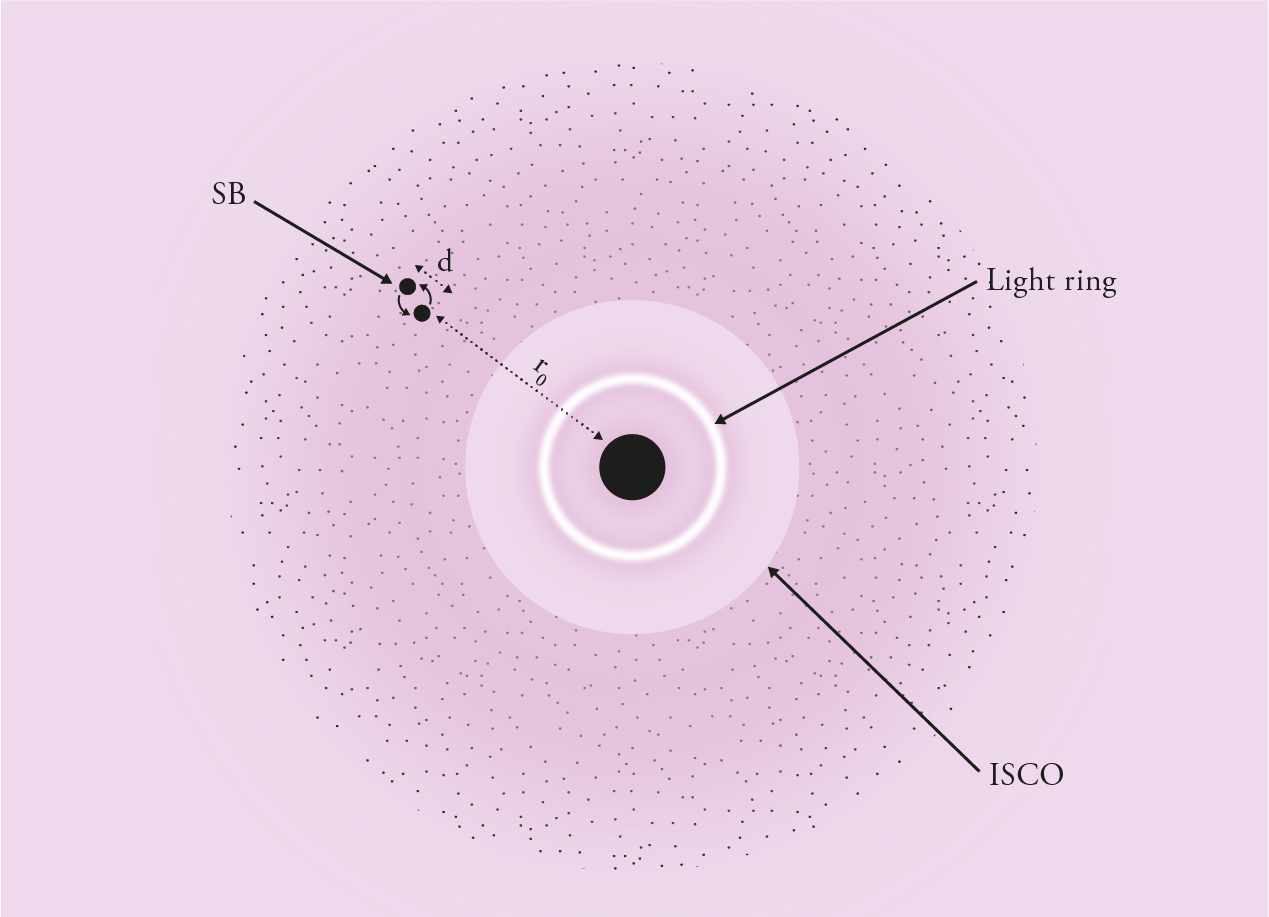}
\caption{\justifying
Depiction of the binary extreme-mass-ratio inspiral (b-EMRI) considered herein.
Adapted from Ref.~\cite{Cardoso:2021vjq}.
A Kerr SMBH sits at the center and acts as the primary of a b-EMRI system.
The ISCO, light ring and horizon of the SMBH all play a role in GW emission from this system.
The secondary binary (SB, not to scale) lies in the vicinity of the SMBH.
The tuning fork is the stellar binary, whose intrinsic frequency varies as the binary evolves, possibly resonating with the modes of the central BH.
Such systems may be ubiquitous in the cosmos.}
\label{fig:b_EMRI}
\end{figure}
In this paper, we study the b-EMRI depicted in Fig.~\ref{fig:b_EMRI}.
The setup consists of an equal-mass binary (hereafter the stellar-mass binary, SB) in the spacetime of a Kerr supermassive BH (SMBH) with mass $M$ and spin parameter $a$.
In Boyer-Lindquist coordinates $\{t,r,\theta,\phi\}$, the CoM of the SB is held fixed in the equatorial plane, $\theta=\pi/2$, at a radius $r=r_0$.
Each component of the SB has mass $m_p$ and executes circular motion about the CoM with separation $d$.
We focus on scenarios exhibiting a sharp hierarchy of scales, $m_p \ll d \ll M \lesssim r_0$.
We allow for a generic inclination of the SB spin with respect to the $\hat{z}$-axis (see Fig.~\ref{fig:BinaryGeometry}), and denote this angle by $\iota_{\rm SB}$. 

In Ref.~\cite{Santos:2025ass}, we implemented a more realistic model in which the CoM of the SB follows a circular orbit around the SMBH.
However, that setup is considerably more intricate and does not readily admit a simple interpretation of the results.
Since our focus here is on the resonant excitation of black-hole modes and their geometric interpretation, we instead consider the idealized case in which the CoM of the SB is at rest with respect to distant observers.
This configuration can be treated in an essentially identical manner and significantly simplifies the analysis by rendering the GW signal monochromatic, as we demonstrate below.

To model the SB in the SMBH spacetime, we use the scale separation $d \ll M$ and apply Dixon's formalism~\cite{Dixon:1970zza,Dixon:1974xoz,Dixon:2015vxa,Costa:2014nta,Harte:2011ku}.
In this approach, the SB is effectively described as a point particle endowed with internal structure up to quadrupolar order~\cite{Steinhoff:2009tk,Harte:2020ols} so as to capture the gravitational radiation generated by its internal motion.
The spin and quadrupole moment of the SB are tensors in the tangent space to the CoM, with explicit forms given in Eq.~(S.30) of the Supplemental material of Ref.~\cite{Santos:2025ass}.

Holding the SB fixed leads to significant simplification: the system is characterized by a single frequency, namely the intrinsic frequency of the SB’s internal motion,
\begin{equation}
    \Omega_{\rm SB} = \frac{1}{u^t}\sqrt{\frac{2 m_p}{d^3}} \, , \label{eq:frequencies_BL}
\end{equation}
where $u^t$ denotes the $t$ component of the four-velocity of the CoM of the SB.
To compute the gravitational perturbations generated by the b-EMRI on the Kerr background, we solve the Teukolsky equation~\cite{Teukolsky:1972my,Teukolsky:1973ha} in the frequency domain.
Imposing purely outgoing boundary conditions at future null infinity, together with purely ingoing boundary conditions at the event horizon, one readily finds the asymptotic form of the Weyl scalar,
\begin{equation}
    \psi_4 \sim \sum_{\substack{\ell, m,q}} {Z}^\infty _{\ell m q} \frac{e^{-i\omega_{q} u}}{r} \, _{-2}S_{\ell m \omega_{q}} (\theta) e^{i m \phi} \, , \label{eq:psi4_inf}
\end{equation}
where $\{u,r,\theta,\phi\}$ are (retarded) Boyer-Lindquist coordinates in the SMBH spacetime, while $_{-2}S_{\ell m \omega}$ denote the $s=-2$ spin-weighted spheroidal harmonics~\cite{Press:1973zz,1986JMP....27.1238L} with spheroidicity $a\omega$, angular mode number $\ell\geq2$, and azimuthal mode number $m$ in the range $-\ell \leq m \leq \ell$.
The frequencies excited by the system are 
\begin{equation}
    \omega_{q} = q \Omega_{\rm SB} \, , \label{eq:frequencies}
\end{equation}
where $q\in\{\pm2\}$ is the quadrupole mode number.
The resulting signal is clearly monochromatic.
The range of $q$ reflects the quadrupole approximation taken for the source.
In contrast with the full model developed in Ref.~\cite{Santos:2025ass}, the frequencies are now only labeled by the quadrupole number precisely because the CoM is held static; otherwise, $\omega$ would also receive contributions from the orbital timescale and geodesic precession~\cite{vandeMeent:2019cam,Rindler_Perlick_1990}.
The amplitudes ${Z}^\infty_{\ell m q}$ take the form
\begin{equation}
    Z_{\ell m q}^\infty = \sum_{i=0}^{4}\ \sum_{j=0}^{4-i}
    \mathcal{A}^{(i,j)}_{\ell m q} \frac{d^i }{dr^i} R^{H}(r_0) \frac{d^j}{d\theta^j}  \bar{S}(\pi/2) \, , \label{eq:amplitudes}
\end{equation} 
where $\mathcal{A}$ depends only on the (both outer and inner) orbital parameters of the SB, $R^{H}\equiv R^{H}_{\ell m \omega_{q}}$ is a solution to the homogeneous Teukolsky equation satisfying purely ingoing boundary conditions at the horizon, and $\bar S \equiv {_{-2}\bar{S}_{\ell m \omega_{q}}}$.
These eigenfunctions are calculated semi-analytically~\cite{Mano:1996vt,Fujita:2004rb,Leaver:1985ax,Berti:2005gp,vandeMeent:2015lxa}.
In a given simulation, for each value of $q$ we must compute the amplitudes $Z_{\ell m q}^\infty$ for many harmonics $(\ell,m)$ until we meet a convergence criterion at $\ell=\ell_{\rm max}\sim r_0\Omega_{\rm SB} $.
This criterion is defined in Eq.~(S.5.B) of the Supplemental material of Ref.~\cite{Santos:2025ass}. 

We can compute the energy flux at infinity using the formulas in~\cite{Teukolsky:1974yv}:
\begin{align}
    \dot{E}^\infty =&  \sum_{\ell}^{\ell_{\rm max} } \dot{E}^\infty_{\ell} \, , \quad  \dot{E}^\infty _{\ell} = \sum_{m,p, q} \frac{|Z^\infty_{\ell m q} |^2}{4 \pi\omega_{q}^2} \, . \label{eq:energy_inf}  
\end{align}
A similar procedure yields the energy flux on the horizon
\begin{align}
    \dot{E}^H =& \sum_{\ell}^{\ell_{\rm max} } \dot{E}^H _{\ell} \, , \quad \dot{E}^H_\ell = \sum_{m,p, q} \alpha_{\ell m q}\frac{|Z^H_{\ell m q} |^2}{4 \pi\omega_{ q}^2} \, , \label{eq:energy_hor}
\end{align}
where the specific expression for $\alpha_{\ell m q}$ can be found in Eq.~(4.44) of Ref.~\cite{Teukolsky:1974yv}.
The amplitudes on the horizon $Z^H_{\ell m q}$ are given by Eq.~\eqref{eq:amplitudes} with $R^H \to R^\infty$, the solution of the homogeneous Teukolsky equation satisfying outgoing boundary conditions at infinity. 

The sum over modes in Eq.~\eqref{eq:psi4_inf} can be made more efficient when the system enjoys a reflection symmetry about the equatorial plane ($\theta=\pi/2$).
In that case, the amplitudes satisfy $Z^{H,\infty}_{\ell (-m)(- q)} =(-1)^\ell\bar{Z}^{H,\infty}_{\ell m q} $. We will make use of this symmetry throughout, unless stated otherwise.
%
%%%%%%%%%%%%%%%%%%%%%%%%%%%%%%%%%%%%%%%%%%%%
\subsection{Static b-EMRIs as tuning forks}
%%%%%%%%%%%%%%%%%%%%%%%%%%%%%%%%%%%%%%%%%%%%
%
\begin{figure}[ht!]
\includegraphics[width=.5 \textwidth]{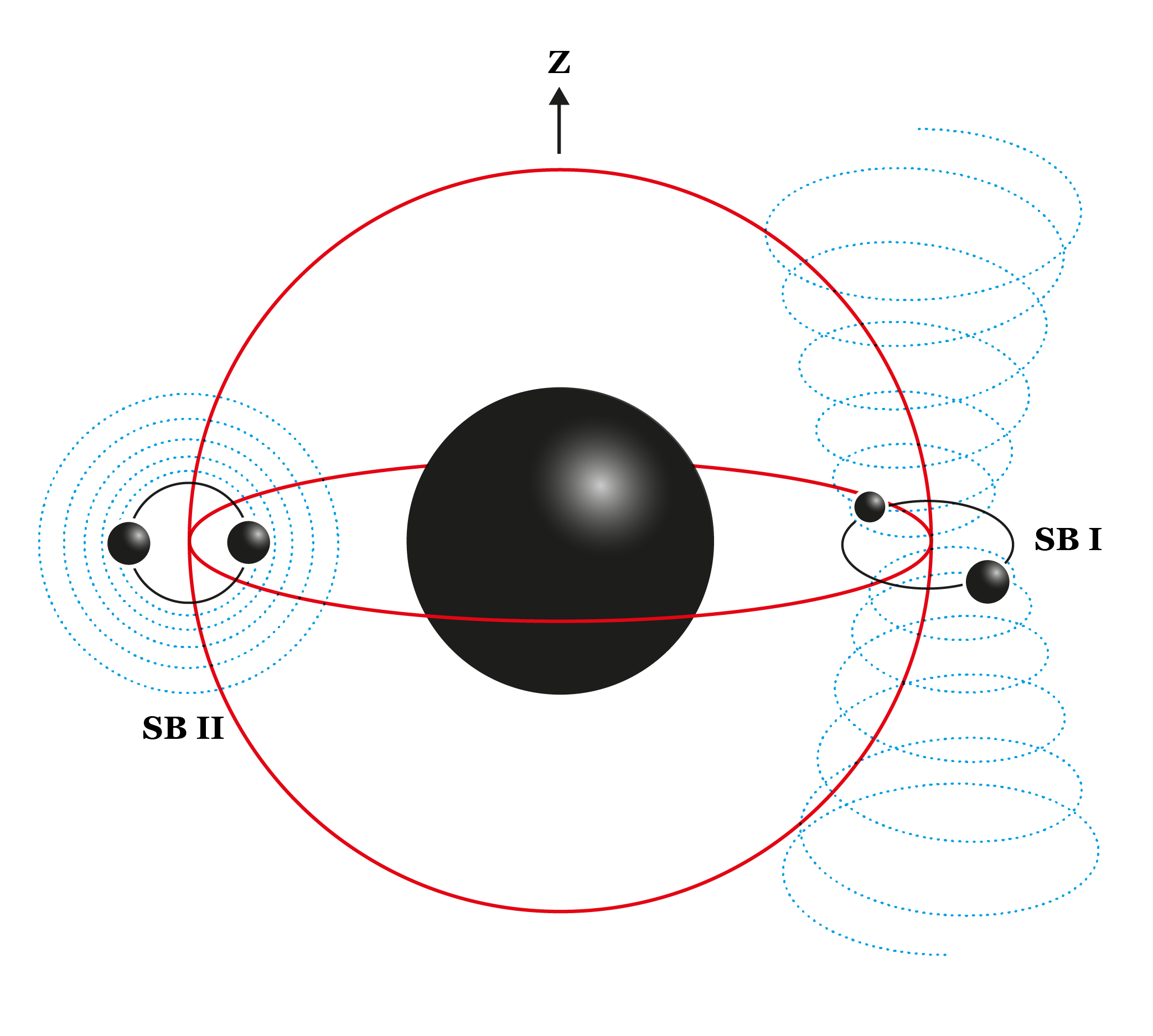}
\caption{\justifying 
The geometrical setup that we consider in Sec.~\ref{sec:LR}.
The primary is a supermassive, non-spinning BH.
A reference SB is placed at the photon sphere, with intrinsic spin parallel to the $\hat{z}$-axis of a reference frame centered at the supermassive BH (SB I), and we also study a configuration with the intrinsic spin aligned with the equatorial light ring (SB II).
As we discuss below, most of the radiation (indicated by spiraling blue lines) is emitted in the direction of the SB intrinsic spin, and so the relative orientation between this spin and the $\hat z$-axis determines the excitation of different angular modes of the SMBH.
The polar and equatorial light rings are depicted by red circles.
Thus, SB I is emitting most of its radiation onto the polar light ring, whereas SB II is sending most of the radiation to the equatorial light ring. 
}
\label{fig:BinaryGeometry}
\end{figure}
As mentioned above, we focus on the relatively simple case of a b-EMRI with a static CoM and investigate its ability to resonantly excite the modes of the SMBH.
Below, we present a series of numerical simulations in which we vary the intrinsic binary frequency $\Omega_{\rm SB}$ while keeping the separation fixed at $d = 0.01M$.
We analyze the dependence of the response on the distance $r_0$ from the SMBH, as well as on the inclination $\iota_{\rm SB}$ of the stellar-mass binary and the SMBH spin parameter $a$.
A diagram showing two of the possible geometries is shown in Fig.~\ref{fig:BinaryGeometry}.

The high-frequency character of QNMs, $M \omega^{\rm QNM}_{\ell m n} \gtrsim \mathcal{O}(1)$, implies that they cannot be efficiently excited by regular EMRIs, whose characteristic frequencies satisfy $M \omega\lesssim 1$.
In contrast, b-EMRIs naturally produce radiation at frequencies $\omega \sim \pm 2 \Omega{\rm SB}$, allowing for $M \Omega_{\rm SB} \gtrsim 1$ and hence efficient excitation of QNMs.
Indeed, previous work has demonstrated that the high-frequency radiation generated by b-EMRIs can resonate with the QNMs of the SMBH~\cite{Cardoso:2021vjq}.
Our goal here is to understand how the BH response is maximized as a function of the SB frequency. The toy model presented in Appendix~\ref{sec:Toy} suggests that the frequency at which the response is maximized does not coincide exactly with the QNM frequencies of the system, a conclusion also supported by earlier numerical studies of a related---albeit more idealized---setup~\cite{Cardoso:2021vjq}.

A meaningful discussion of the results requires a consistent normalization of the energy fluxes.
Throughout this work, we will often normalize the energy fluxes to the redshifted quadrupole formula~\cite{Santos:2025ass} 
\begin{eqnarray}
 & \dot E^{Q} &= \frac{8}{5} (u^t)^4 m_p^2 d^4 \Omega_{\rm SB}^6 \, .\label{eq:EQ}
\end{eqnarray}
This expression provides a good approximation only in the high-frequency regime and is not expected to be accurate when $2M\Omega_{\rm SB}\lesssim 1$.
%
%
%%%%%%%%%%%%%%%%%%%%%%%%%%%%%%%%%%%%%%%%%%%%
\section{Results} \label{sec:Results}
%%%%%%%%%%%%%%%%%%%%%%%%%%%%%%%%%%%%%%%%%%%%

%%%%%%%%%%%%%%%%%%%%%%%%%%%%%%%%%%%%%%%%%%%%%%%%%%%%%%%%%%%%%%%%%%%%%%%%%%%%%%
\subsection{An off-center binary in flat space} \label{sec:Flat}
%%%%%%%%%%%%%%%%%%%%%%%%%%%%%%%%%%%%%%%%%%%%%%%%%%%%%%%%%%%%%%%%%%%%%%%%%%%%%%
%
Before tackling the more complicated system described in the previous section, we comment on a much simpler one: that of a Newtonian binary sitting in a \emph{generic point} in flat space. If the binary is at the origin of the coordinate system and moves on the $xy$-plane, it almost exclusively excites the $(\ell,m) = (2,\pm2)$ modes~\cite{Peters:1963ux}. However, if the binary is moved from the origin, it will excite a spectrum of modes. The problem at hand is equivalent to finding how spin weighted spherical harmonics on the celestial sphere transform under a translation of the origin. The solution will be a generalization of the perturbative result obtained in Ref.~\cite{Boyle:2015nqa}. Moreover, due to the symmetries of flat space, we can assume that the binary is sitting on the $\hat x$-axis at a distance $r_0$ from the origin. We then allow its spin to have a generic orientation with respect to the coordinate axes. 

First, we point out that, in this case, we only care about the asymptotic behavior of the field at infinity, as there is no flux through a BH horizon. The framework is otherwise similar, albeit significantly simpler, to the one used in curved space. We outline it in Appendix~\ref{app:FlatFormalism}. To emphasize that the calculation is now in flat space, we denote the amplitudes at infinity by $Z^{\rm N} _{\ell m q}$. The associated energy fluxes are calculated using 
\begin{align}
    \dot{E}^{\rm N} =&  \sum_{\ell}^{\ell_{\rm max} } \dot{E}^{\rm N}_{\ell} \, , \quad  \dot{E}^{\rm N} _{\ell} = 2 \sum_{m}  \dot{E}^{\rm N} _{\ell m 2} = \sum_{m} \frac{|Z^{\rm N}_{\ell m 2} |^2}{8 \pi\Omega_{\rm SB}^2} \, , \label{eq:energy_inf_flat}  
\end{align}
where $\dot{E}^{\rm N} _{\ell m 2}$ is the energy flux in a given $(\ell,m)$ mode for $q=2$. The factor of 2 before the sum over $m$ accounts for the modes with $q=-2$, leveraging the symmetry in the amplitudes $Z^{\rm N}_{\ell(-m)(-q)} = (-1)^\ell Z^{\rm N}_{\ell mq}$. As a first sanity check, we obtained
\begin{equation}
    \frac{|\dot E^{\rm N}- \dot E^Q|}{\dot E^{\rm N}} < 10^{-12}, \label{eq:benchmark_flat}
\end{equation}
which is consistent with the precision goal of the convergence criterion, which aims for at least five digits in the amplitude. 

\begin{figure}[t]
\centering
\includegraphics[width=.45 \textwidth]{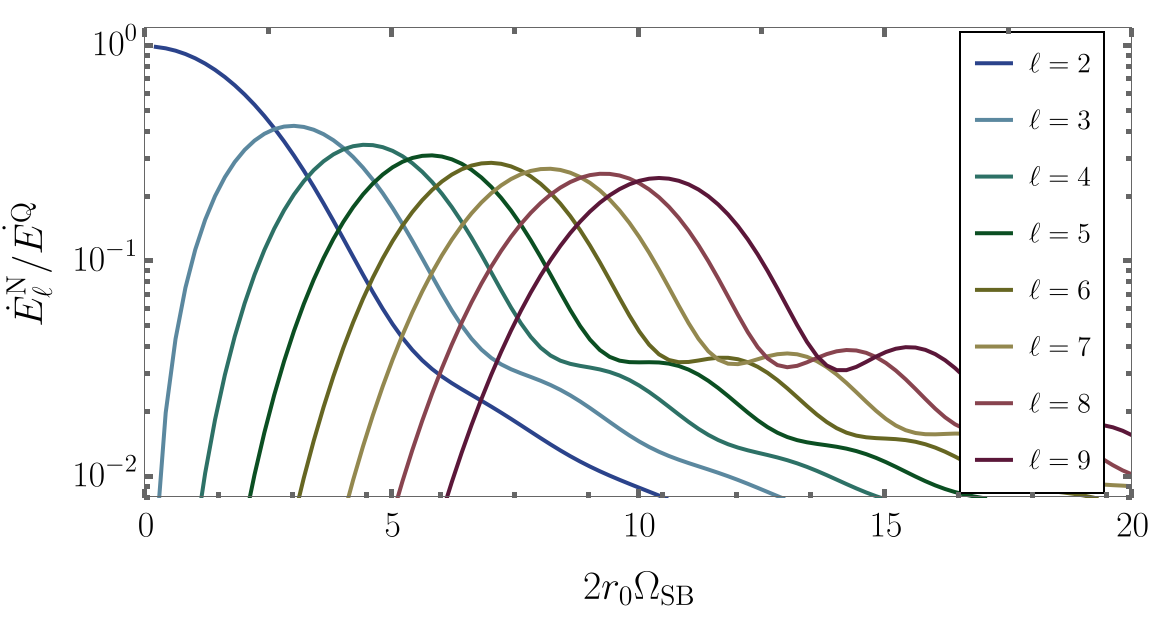}
\caption{\justifying
Energy flux in different $\ell$ modes, normalized to total flux $\dot E ^Q$, for a Newtonian binary system with orbital frequency $\Omega_{\rm SB}$ in flat space, with the CoM at $x=r_0=10$ (in arbitrary units), $y=z=0$.
The spin of the binary points in the $\hat{z}$ direction; we use the quadrupole approximation, so that the GW frequency is simply $2\Omega_{SB}$.
Still, the signal is not uniquely described by $\ell=m=2$ harmonics, as would be the case if the CoM were at the origin.
Each curve has a global maximum for $2 r_0 \Omega_{\rm SB} \approx 1.1 \sqrt{\lambda_\ell}$, where $\lambda_\ell= \ell(\ell+1)-2$ is the eigenvalue of the spin $-2$ harmonics with angular number $\ell$.}
\label{fig:FlatModes}
\end{figure}

In Fig.~\ref{fig:FlatModes}, we show the energy flux in the first few $\ell$ modes normalized to the total flux, for a binary with spin pointing in the $\hat{z}$ direction. It is clear that the $\ell=2$ modes are not dominant for large $r_0\Omega_{\rm SB}$. Indeed, we find that each $\ell$ mode has a maximum in the energy flux when $\omega \equiv 2\Omega_{\rm SB} = \omega^{\rm geom}_{\ell}$, for 
\begin{equation}
    \omega^{\rm geom}_{\ell}\approx  1.1  \frac{\sqrt{\lambda_{\ell}}}{r_0} \, , \label{eq:geometric_freq}
\end{equation}
where $\lambda_\ell= \ell (\ell+1) -2$ is the eigenvalue of spin $-2$ spherical harmonics with angular number $\ell$, and the numerical coefficient is found empirically. We call this frequency $\omega^{\rm geom}$ because it is not a special frequency in any way, and the peak in energy flux is entirely due to the geometry of the problem. Recall that in this case the GW signal is just that of a textbook Newtonian binary in the quadrupole approximation; we are simply choosing a bad frame to describe it. 

The amplitude for a generic orientation of the binary and harmonic mode is very complicated and would not fit in a page. For the case where the spin of the binary is pointing in the $\hat{z}$ direction, the amplitudes for the $(\ell,m)=(2,2)$ and $(3,3)$ modes are ($\eta_0 \equiv 2 \Omega_{\rm SB} r_0$)
\begin{align}
    &\frac{Z^{\rm N}_{222}}{m_p d^2}=-\frac{\sqrt{5 \pi } }{32 \,
   r_0^4} \eta_0^2
   \left(\left(\eta_0^2-48\right) j_2(\eta_0)+9 \eta_0 \, j_3(\eta_0)\right) \, , \\ 
   &\frac{Z^{\rm N}_{332}}{m_p d^2}=\frac{\sqrt{42 \pi } }{64 i \, r_0^4}
   \eta_0^2 \left(\left(\eta_0^2-80\right) j_3(\eta_0)+11 \eta_0\, 
   j_4(\eta_0)\right) \, ,
\end{align}
where the $j_\ell$ are spherical Bessel functions of the first kind. It is easy to check that for sufficiently large $\eta_0$ the $(2,2)$ mode ceases to be dominant, consistent with Fig.~\ref{fig:FlatModes}. To illustrate this point further, we can take the leading terms in a small $\eta_0$ expansion for the corresponding energy fluxes:
\begin{equation}
    \frac{2 \dot E^{\rm N}_{222}}{\dot E^{\rm Q}}\sim 1 - \frac{5 \eta_0^2}{21} \, , \quad  \frac{2 \dot E^{\rm N}_{332}}{\dot E^{\rm Q}}\sim  \frac{5 \eta_0^2}{42} \, . 
\end{equation}
Indeed, we find that the only modes that are non-zero for $\eta_0=0$ are the $(2,\pm2)$ modes. As $\eta_0$ increases, we need higher order modes to recover the total flux $\dot E^{\rm Q}$. 

The takeaway message from this flat space analysis is that a non-monotonic profile of excitation of different harmonics is not indicative of any particularly exciting physics, as it may be caused by the choice of BMS frame at infinity~\cite{Bondi:1962px,Sachs:1962wk,Newman:1966ub,Boyle:2015nqa}. 
%
%
%%%%%%%%%%%%%%%%%%%%%%%%%%%%%%%%%%%%%%%%%%%%%%%%%%%%%%%%%%%%%%%%%%%%%%%%%%%%%%
\subsection{Binary at light ring of supermassive BH} \label{sec:LR}
%%%%%%%%%%%%%%%%%%%%%%%%%%%%%%%%%%%%%%%%%%%%%%%%%%%%%%%%%%%%%%%%%%%%%%%%%%%%%%
%
In the high-frequency regime, the QNMs of the BHs are associated with null particles trapped around the BH~\cite{Cardoso:2008bp,Dolan:2010wr,Yang:2012he}. For non-spinning BHs, we therefore expect that a SB located {\it at} the light ring will excite QNMs the most, in the same manner that piano strings are best hit at anti-nodes. We thus start our discussion with this particular location.
%
%%%%%%%%%%%%%%%%%%%%%%%%%%%%%%%%%%%%%%%%%%%%
\subsubsection{Total energy flux}
%%%%%%%%%%%%%%%%%%%%%%%%%%%%%%%%%%%%%%%%%%%%
%
%
\begin{figure}[ht!]
\centering
\includegraphics[width=.45 \textwidth]{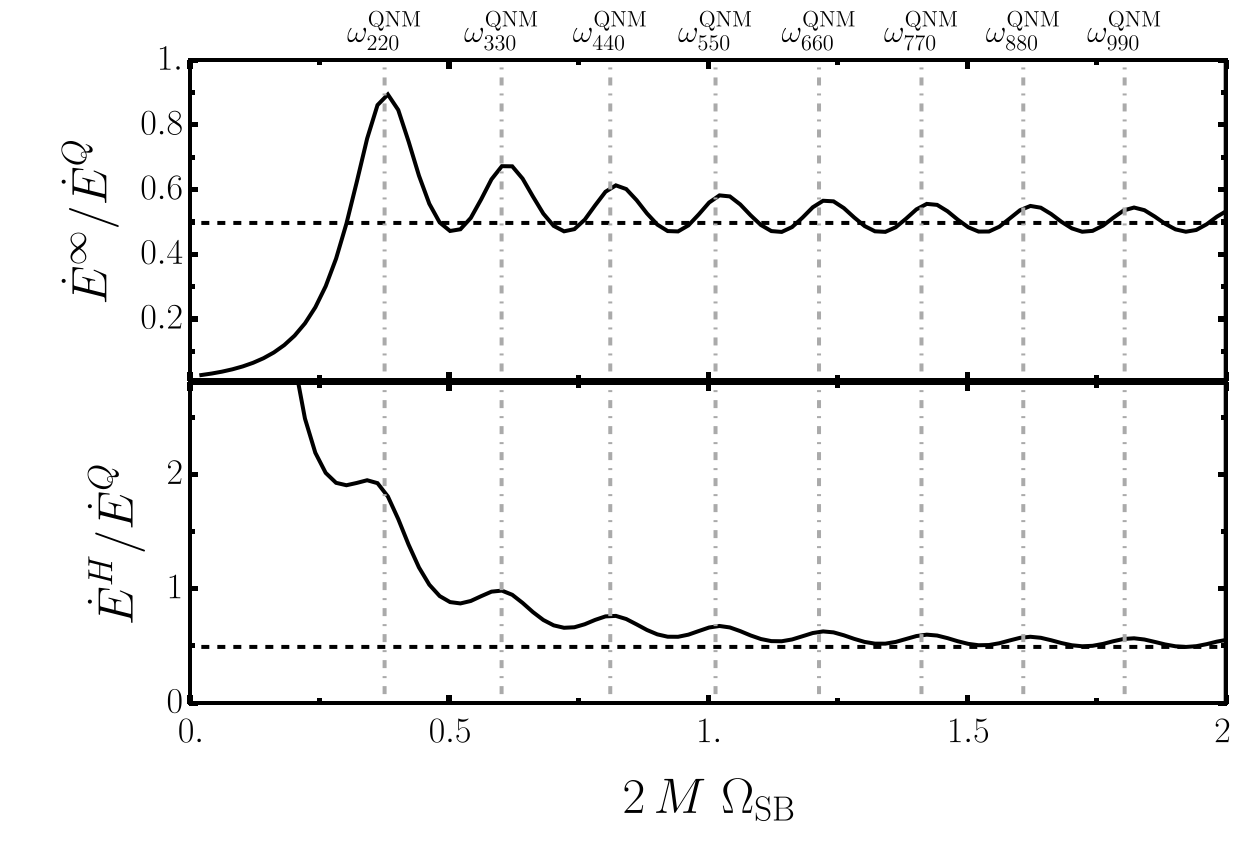}
\caption{\justifying
Total energy flux from SB I (cf. Fig.~\ref{fig:BinaryGeometry}) around a nonspinning BH, at infinity ({\bf top}) and at the horizon (\textbf{bottom}), normalized to the geometric-optics expression $\dot{E}^Q$ in Eq.~\eqref{eq:EQ}.
The SB is static and placed at $r_0=3M$.
Vertical dashed lines indicate the fundamental QNM frequencies $\omega_{\ell m 0}^{\rm QNM}$ for $\ell=m=2,\ldots,9$.
The gravitational radiation is monochromatic with frequency $\omega=2\Omega_{\rm SB}$.
For $M\omega \gg 1$, fluxes at infinity and the horizon asymptote to roughly $\dot{E}^Q/2$, as expected, since at the light ring half the gravitons escape to infinity whereas the other half fall into the BH.
For small $M \omega$, $\dot{E}^Q$ is outside its regime of validity, but we checked that both $\dot E^{\infty,H}\to 0$, while $\dot E^{H}/\dot E ^{\infty}\to \infty$ when $\omega\to 0$; the latter is an instance of the horizon dominance effect first reported in Ref.~\cite{Santos:2024tlt}.
The resonances with the QNMs lead to an excess in the energy flux, easily identified above.}
\label{fig:Sch_LR_total}
\end{figure}
Fig.~\ref{fig:Sch_LR_total} shows total fluxes at infinity and at the horizon, when the SB is held static at the light ring of a Schwarzschild BH, $r_0 = 3M$, and with its intrinsic spin parallel to the $\hat{z}$-axis of a frame centered at the SMBH (SBI in Fig.~\ref{fig:BinaryGeometry}). The vertical dashed lines indicate the real parts of the first few fundamental QNM frequencies (the least damped modes, $n=0$, are expected to exhibit the strongest excitations).

The resonances with the BH QNMs lead to an increase in the energy fluxes at certain discrete SB frequencies, a feature that is apparent in the figure. The resonant frequencies are very close to, but not exactly at, the BH QNM frequencies. This slightly puzzling feature had been seen before~\cite{Cardoso:2021vjq,Yin:2024nyz}, and we show in Appendix~\ref{sec:Toy} a very simple toy model of a dissipative system with the same behavior. This misalignment has also been known for decades in other systems~\cite{Purcell:1946zz,PhysRevLett.110.237401}, and, more recently, it was found for resonances between a binary system and a boson cloud in Ref.~\cite{Brito:2023pyl}. We will explore it further in the rest of this work, but clearly an in-depth, first-principles treatment in BH spacetimes is missing in the literature.

At high frequencies, both total energy fluxes oscillate around $\dot E^{\infty , H} \approx \dot E ^Q / 2$. Indeed, $\dot E ^Q$ gives a good approximation for the total energy flux, but the strong lensing at the light ring causes half of the radiation to escape to infinity, while half is captured at the horizon~\cite{Barausse:2021xes}.

At low frequencies, $\dot E^Q$ is not a meaningful estimate of the total flux of the SB, but Fig.~\ref{fig:Sch_LR_total} still gives us interesting information on how $\dot E^{\infty , H}$ behave for small $\Omega_{\rm SB}$ compared to $\dot{E}^Q \sim \Omega_{\rm SB}^6$. We find the following behavior in the low frequency regime:
\begin{equation}
    \dot{E}^{\infty} \sim \Omega_{\rm SB}^6 \, , \quad  \dot{E}^{H} \sim \Omega_{\rm SB}^{2} \quad (\Omega_{\rm SB} \to 0) \, . 
\end{equation}
This is a counterintuitive result: at low frequencies, the radiation should barely be able to see a BH of size $M$. Nevertheless, most of the radiation is going into the horizon, rather than being scattered to large distances. This behavior has been seen in other setups, and it has been dubbed the ``horizon dominance effect"~\cite{Santos:2024tlt,Santos:2024okf}. This is a potentially interesting property of b-EMRIs, as it might lead -- when the central BH spins and can thus power superradiance~\cite{Brito:2015oca} -- to overall energy extraction from the BH being deposited into the binary, perhaps stalling the inspiral and leading to floating orbits~\cite{Press:1972zz,Cardoso:2011xi}. We will not explore this further here.

%
%%%%%%%%%%%%%%%%%%%%%%%%%%%%%%%%%%%%%%%%%%%%%%%%%%%%%%%%%%%%%%%%%%%%%%%%%%%%%%%%%%%%%%
\subsubsection{Excitation of different \texorpdfstring{$\ell$}{l} modes and a geometric interpretation}
%%%%%%%%%%%%%%%%%%%%%%%%%%%%%%%%%%%%%%%%%%%%%%%%%%%%%%%%%%%%%%%%%%%%%%%%%%%%%%%%%%%%%%
%
\begin{figure}[ht!]
%\centering
\includegraphics[width=.5 \textwidth]{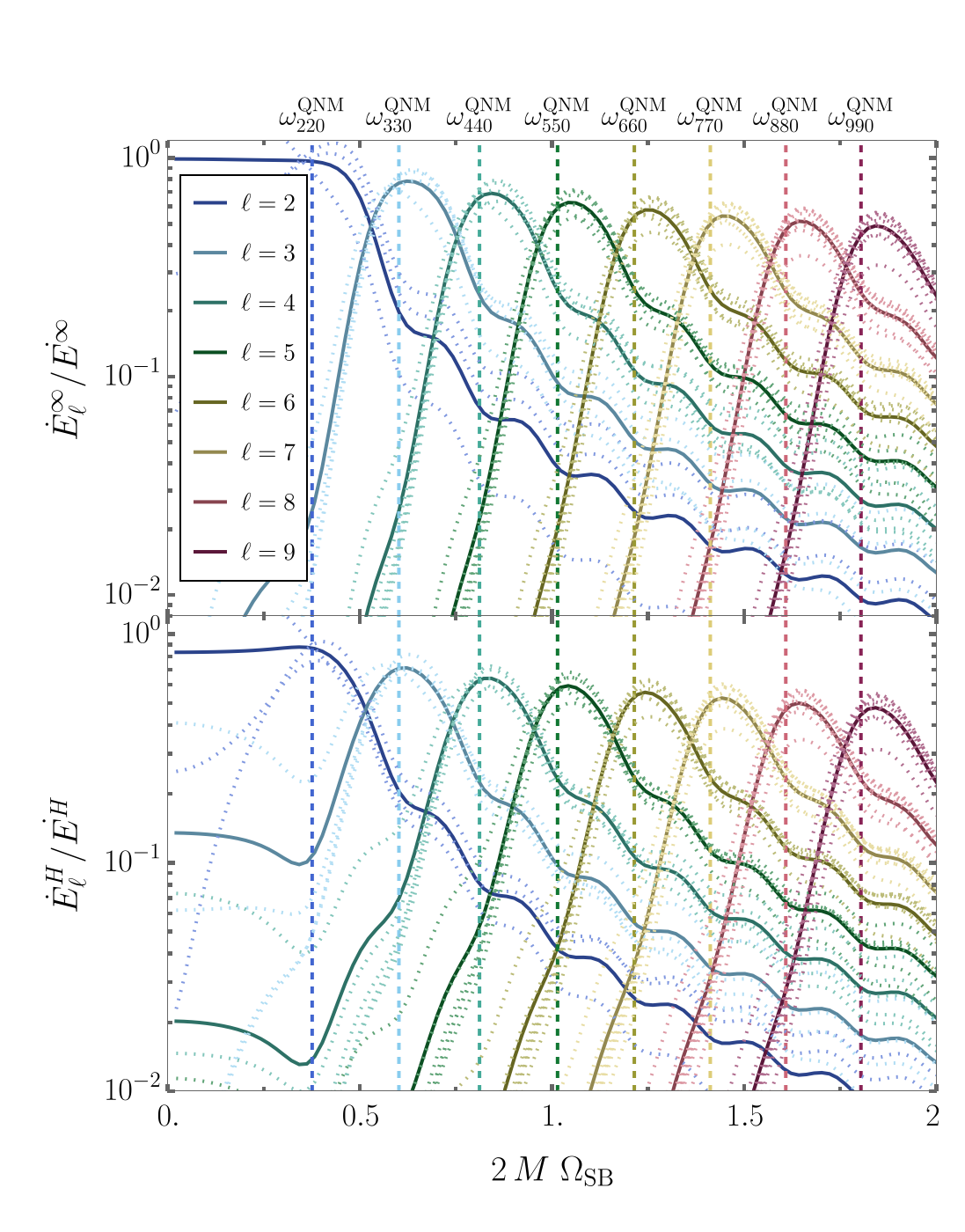}
\caption{\justifying
Energy flux from the binary in Fig.~\ref{fig:Sch_LR_total} for different $\ell$ modes, normalized to the total energy flux at infinity ({\bf top}) and on the horizon ({\bf bottom}), respectively. Different colors indicate different $\ell$ modes.
Vertical dashed lines mark the real part of the QNM frequencies, while the dotted lines are the individual $\dot E_{\ell m}$ (rescaled by $2 \ell +1$) contributing to $\dot E_\ell$.
Each $\ell$ mode peaks when it crosses the corresponding QNM frequency.
The wiggles for large $\Omega_{\rm SB}$ are ``anti-resonances'' caused by peaks in the denominator; see Fig.~\ref{fig:Sch_LR_total}.}
\label{fig:Sch_LR_l_modes}
\end{figure}
\begin{figure}[ht!]
\centering
\includegraphics[width=.45 \textwidth]{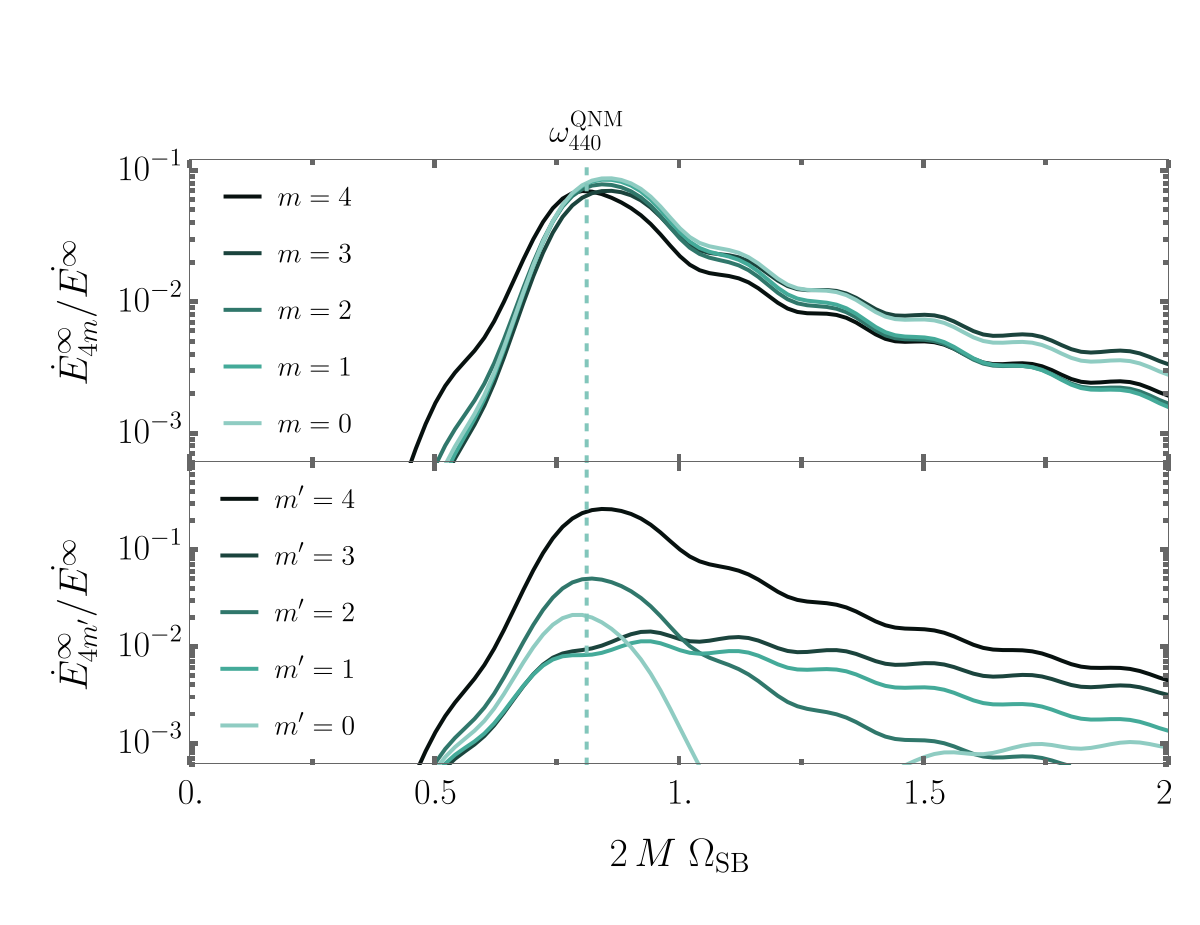}
\caption{\justifying
Energy flux at infinity for different azimuthal modes with $\ell=4$; see Eq.~\eqref{eq:energy_inf}.
The b-EMRI configuration is the same as in Fig.~\ref{fig:Sch_LR_total}, except for the orientation of the SB spin, for which we have on {\bf Top:} the SB spin is tangent to a polar light ring, SB I in Fig.~\ref{fig:BinaryGeometry}; {\bf Bottom:} the SB spin is aligned with the equatorial light ring, SB II in Fig.~\ref{fig:BinaryGeometry}.
In the bottom panel, we see that the $\ell = m^\prime = 4$ mode is particularly excited in this geometric setup, as explained by the discussion in the text.
}
\label{fig:Sch_LR_l=4}
\end{figure}
We have established that resonances are being excited; we now want to understand which modes are resonating with the SMBH. For instance, if we are crossing a resonance associated with $\ell = 2$ modes, we expect the surplus energy flux to also come in modes with $\ell = 2$. To test this hypothesis, and to understand the finer details of the resonance process, we study the energy flux in each $\ell$ mode, normalized to the total flux, in Fig.~\ref{fig:Sch_LR_l_modes}. 

Indeed, as is apparent from the figure, resonances do roughly coincide with an excess of energy flux in the modes with the same value of $\ell$ as the excited QNM frequency. As we remarked, the energy flux is not maximal exactly at the real part of the QNM frequencies. However, in light of the results presented in Sec.~\ref{sec:Flat}, we must be careful not to mistake this correlation for causation. Indeed, as will become clear as the analysis is deepened, the excess energy flux in each $\ell$ mode is mostly due to the geometry of the problem, not to QNM excitation.

Note that the ``wiggles'' in each $\ell$ curve after the global maximum are not resonances in each $\ell$ mode; instead, they are ``anti-resonances" caused by a resonance in the denominator (the total flux) as a QNM frequency is crossed.

Dotted lines in Fig.~\ref{fig:Sch_LR_l_modes} correspond to individual $(\ell, m)$ modes contributing to each solid curve (rescaled by a factor $2 \ell +1$ so they line up nicely). They show that for each $\ell$, all the individual $m$ modes are being excited to a similar extent. This is made explicit in the top panel of Fig.~\ref{fig:Sch_LR_l=4}, where we show the excitation of different $m$ modes for $\ell=4$. To understand this, we recall that the radiation from a binary is preferentially emitted along the intrinsic spin axis of the SB~\cite{Peters:1963ux}, see also Fig.~\ref{fig:BinaryGeometry}.
Thus, in the current setup, where the SB intrinsic spin is tangent to a meridian of SMBH's light sphere (see Fig.~\ref{fig:BinaryGeometry}), most of the radiation is being emitted with high inclinations relative to the equatorial plane. In terms of harmonics, this translates into a comparable excitation of the different $m$ modes for each $\ell$.

Conversely, if we rotate the SB such that its spin now points along the equatorial light ring, as SBII in Fig.~\ref{fig:BinaryGeometry}, wavefronts travel on equatorial geodesics, and we expect the most excited harmonics to be the $\ell=m$ modes. To test this, we can simply rotate our reference frame $(\theta,\phi)\to(\theta^\prime, \phi^\prime)$ such that in the rotated frame the binary spin is in the equatorial plane. For each $\ell$, we must then project the amplitudes of the $m$ mode into $m^\prime$ using Wigner's $\mathcal{D}^\ell_{mm\prime}$ matrices, as described in~\cite{Gualtieri:2008ux,Boyle:2015nqa}.
The results, shown in Fig.~\ref{fig:Sch_LR_l=4} for $\ell=4$, fully support our geometrical interpretation\footnote{In terms of QNM excitation, we would say the GWs are trapped orbiting the SMBH near the maximum of the potential. Then, in the first scenario, a GW traveling on a polar geodesic must be described by a combination of $m$ modes, simply because individual $m$ modes can only describe waves rotating around the $\hat{z}$-axis. Conversely, in the second case, the GWs are mostly trapped on the equatorial light ring, so they are well described by the $\ell=m$ harmonic.}.

%%%%%%%%%%%%%%%%%%%%%%%%%%%%%%%%%%%%%%%%%%%%
\subsection{Varying the distance from the supermassive BH}
%%%%%%%%%%%%%%%%%%%%%%%%%%%%%%%%%%%%%%%%%%%%
%
%
\begin{figure}[t]
\centering
\includegraphics[width=.45 \textwidth]{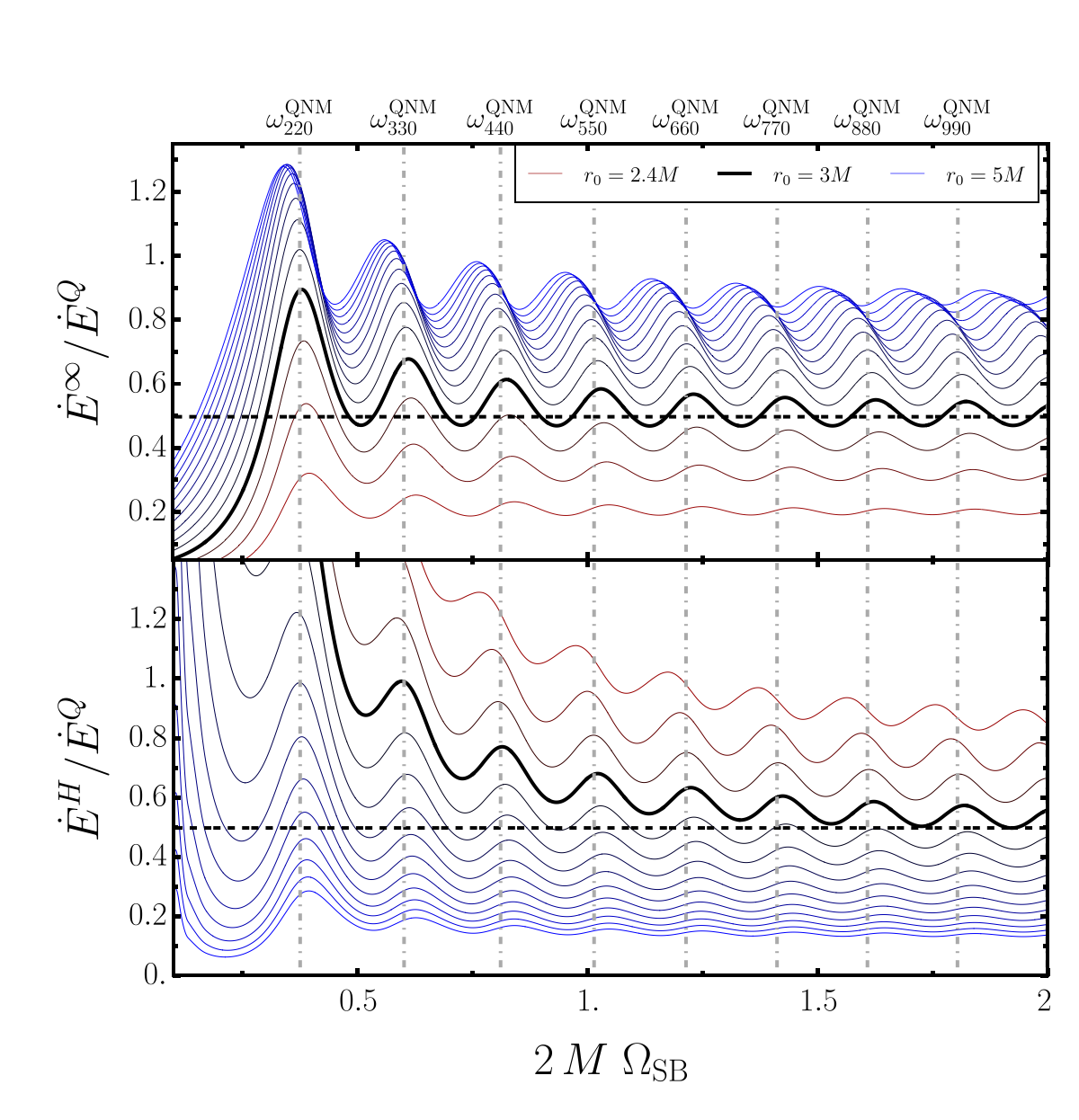}
\caption{\justifying
Same as Fig.~\ref{fig:Sch_LR_total} but for different locations $r_0\in(2.4M,5M)$ of the SB CoM, with increments of $0.2M$.
Varying $r_0$ shifts the frequency at which the energy fluxes peak, as in the toy model presented in Appendix~\ref{sec:Toy}.
The frequencies for which the energy flux at infinity has a maximum decrease as the SB is moved away from the BH; the opposite is true for the maxima of the energy flux at the horizon.}
\label{fig:Sch_multi_total}
\end{figure}
Next, we vary the distance of the SB to the supermassive BH. In Fig.~\ref{fig:Sch_multi_total} we consider the same b-EMRI setup as before, but now placing the SB at  $r_0\in(2.4M,5M)$, with increments of $0.2M$. The energy fluxes have a maximum {\it close to, but not at} the real part of each QNM frequency. We define $\omega_{\ell m} ^{\rm max}$ to be the frequency at which the flux to infinity has a maximum close to ${\rm Re}[\omega^{\rm QNM}_{\ell m 0}]$. Alternatively, it is the frequency corresponding to the peak in $\dot E ^\infty$ which migrates away from ${\rm Re}[\omega^{\rm QNM}_{\ell m 0}]$ as the SB is moved away from the light ring. Of course, for Schwarzschild geometries the azimuthal index is redundant.

As we try to convey with the toy model in Appendix~\ref{sec:Toy}, the response of a dissipative system close to resonance is nontrivial. Even in such a simple toy model, the dependence of the response on the location of the excitation is highly complex.
For our BH-SB system, $r_0$ plays the role of $x_0$, and it is highly nontrivial to predict the behavior of $\omega_{\ell m} ^{\rm max}(r_0)$. Empirically, we find that if the SB is moved further from the BH, $\omega_\ell ^{\rm max}$ tends to decrease, asymptotically approaching $\omega_{\ell m} ^{\rm max}(r_0) \sim k\, {\rm Re}[\omega_{\ell m 0} ^{\rm QNM}]/r_0$, with $k\sim 3M $. For the horizon fluxes, we find the opposite trend, with $\omega_\ell ^{\rm max}$ increasing with $r_0$; this likely results from the approximate symmetry of the potential under reflections about the light ring. 

\begin{figure}[t]
\centering
\includegraphics[width=.49 \textwidth]{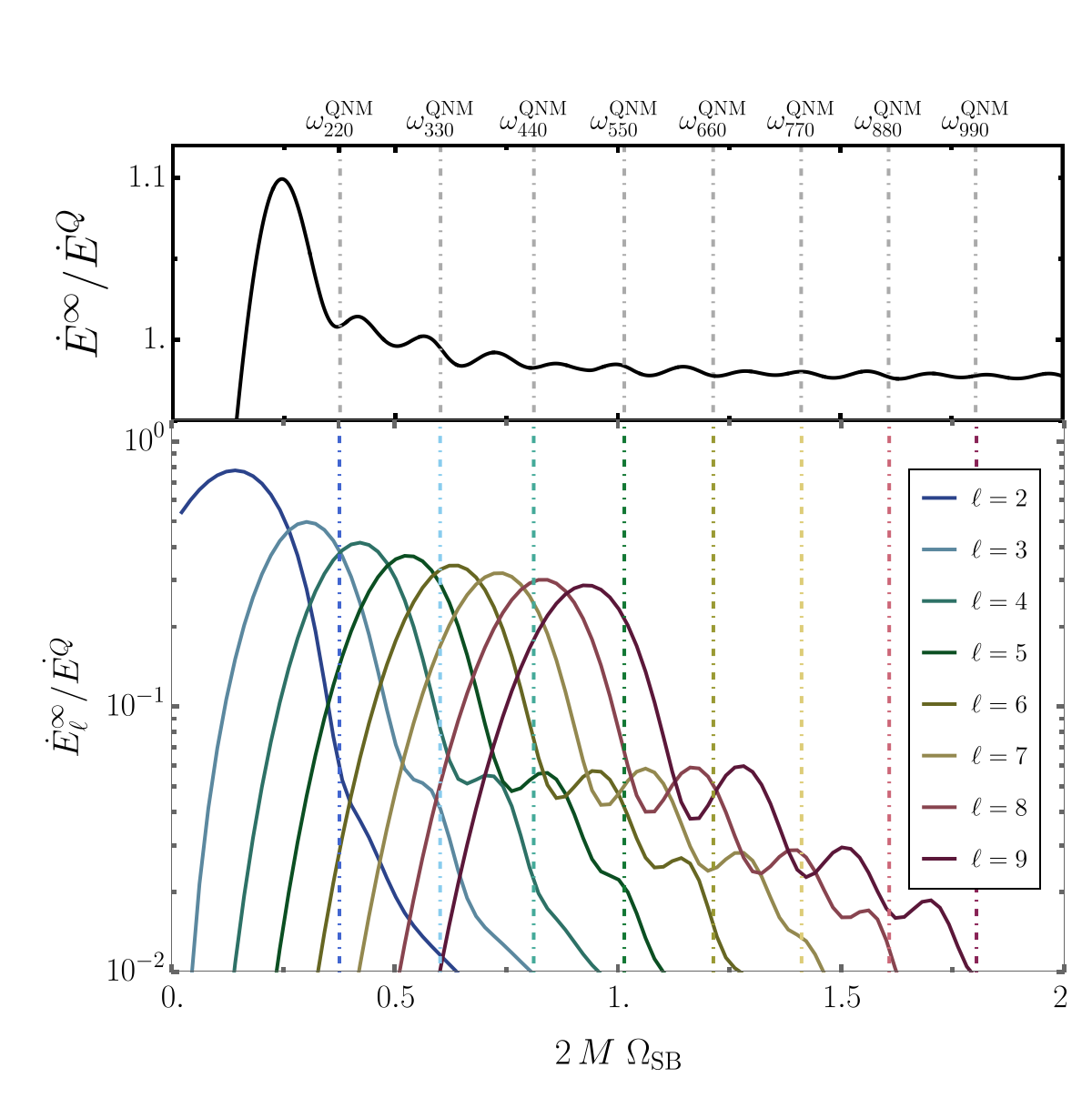}
\caption{\justifying
{\bf Top}: Total energy flux at infinity; {\bf Bottom}: Energy flux at infinity for individual $\ell$ modes, normalized to $\dot E^Q$.
Results for a b-EMRI identical to that of Fig.~\ref{fig:Sch_LR_total} but at $r_0=10M$.
The peaks in the flux of individual modes do not, in general, correspond to any peak in the total flux.
Indeed, the structure of excitation of individual modes is extremely similar to that of flat space (see Fig.~\ref{fig:FlatModes}), where there are no QNMs to excite.
We conclude that only the total flux can be used as a smoking gun for excitation of modes of the SMBH.}
\label{fig:Sch_r=10}
\end{figure}

The shift of the resonance away from the real part of the QNM frequencies can be a very noticeable effect. To further illustrate this point, we look in Fig.~\ref{fig:Sch_r=10} at the energy flux at infinity (top panel) for a SB identical to the ones described before but at $r_0=10M$. It is now clear that the peaks in the flux appear to be unrelated to the QNM frequencies. Moreover, even if one looks at the excitation of individual harmonics through $\dot E^\infty _\ell$ (bottom panel), there is still no clear pattern related to the characteristic frequencies of the SMBH. 

Given the results presented in Figs.~\ref{fig:FlatModes} and~\ref{fig:Sch_LR_l_modes}, we hardly expected $\dot E^\infty _\ell$ to be a good smoking gun for resonant excitation. Indeed, the similarities between the bottom panel of Fig.~\ref{fig:Sch_r=10} and Fig.~\ref{fig:FlatModes} are striking, with each curve having a predominant peak followed by smaller peaks in the tail. Moreover, for sufficiently large $r_0$, the dominant peak occurs at a frequency given by Eq.~\eqref{eq:geometric_freq}; this establishes that it is caused by the geometry of the harmonic basis, not by the resonant excitation of SMBH modes. We conclude that \emph{strong excitation of a given harmonic is not indicative of a resonance with that mode of the SMBH}. Instead, one must rely on the total fluxes at infinity and on the horizon, which are frame independent.
%
%%%%%%%%%%%%%%%%%%%%%%%%%%%%%%%%%%%%%%%%%%%%
\subsection{An inclined binary} \label{sec:Inclined}
%%%%%%%%%%%%%%%%%%%%%%%%%%%%%%%%%%%%%%%%%%%%
%
We have established that the SB is capable of exciting the quasinormal modes of the SMBH. We saw that this excitation is clearest when the binary is close to the light ring, as it directly feeds the region where the waves get trapped. Here we examine another possibility to enhance mode excitation, exploiting the fact that the SB emits most of the energy along the axis of its spin. Thus, even if the binary is not at the light ring, whenever its spin is aligned with a critical null geodesic the waves should get trapped in the light ring (at least in the geometric optics approximation). 
\begin{figure}[t]
\centering
\includegraphics[width=.48 \textwidth]{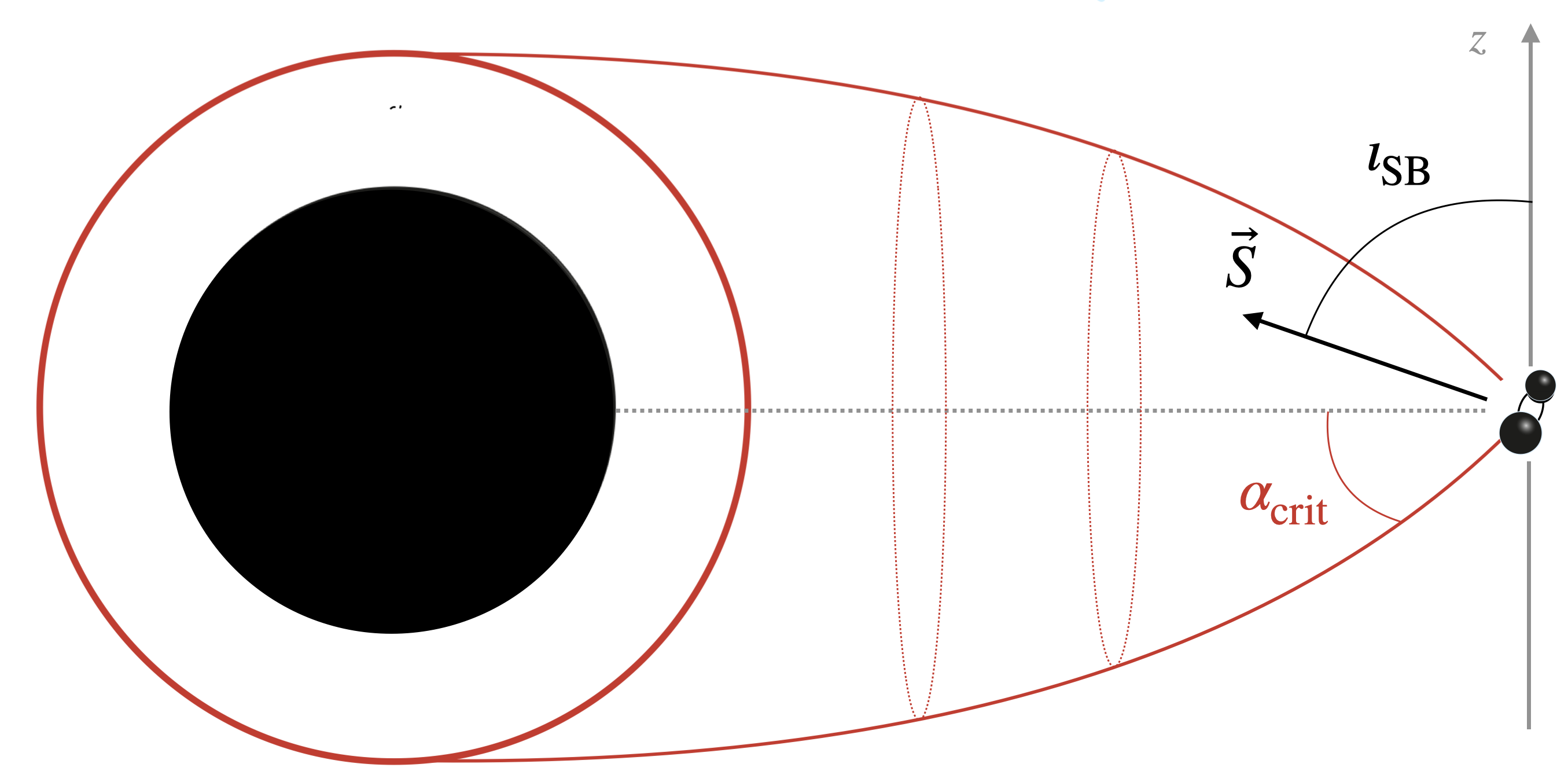}
\caption{\justifying
Geometric setup for the study of an inclined SB (right) at a distance $r_0$ from a SMBH (left).
The SMBH is encircled by a photon sphere, drawn as a thick red ring.
The intrinsic spin $\vec{S}$ of the SB (black arrow) makes an angle $\iota_{\rm SB}$ with the $\hat z$-axis (gray arrow), and lies in the plane containing the $\hat{z}$-axis and the line connecting the SB to the SMBH (dashed gray line).
From the position of the SB, there is a cone of critical null geodesics that asymptotically connect to the light sphere (thin red lines).
The aperture of the cone is $\alpha_{\rm crit}$, and is given in Eq.~\eqref{eq:alpha_c}.}
\label{fig:InclinedGeometry}
\end{figure}

Following the above discussion, we now focus on systems where the intrinsic spin of the SB is at an angle $\iota_{\rm SB}$ with respect to the $\hat z$-axis. The intrinsic spin vector lies in the plane that contains the $\hat z$-axis and the line that connects the SB to the SMBH, as depicted in Fig.~\ref{fig:InclinedGeometry}. Drawing from geometric optics intuition, to maximize QNM excitation we want to maximize the amount of radiation that feeds into critical null geodesics. At the SB position, there is a cone of such geodesics (red in Fig.~\ref{fig:InclinedGeometry}), whose aperture is
\begin{equation}
    \alpha_{\rm crit} = \arcsin \frac{3 \sqrt 3 M}{r_0} \sqrt{1-\frac{2 M }{r_0}} \, . \label{eq:alpha_c}
\end{equation}

If the SB emitted isotropically, changing $\iota_{\rm SB}$ would not change the amount of QNM excitation. If it emitted only along the axis, it would be maximal when $\iota_{\rm SB} = \iota_{\rm crit} = \pi/2 - \alpha_{\rm crit}$. However, neither is the case; in its local rest frame, the SB emits radiation according to the Peters-Matthews formula~\cite{Peters:1963ux}, so it is not trivial to predict which inclination $\iota_{\rm max}$ maximizes the response of the system. Using an analytical estimate outlined in Appendix~\ref{app:Inclined}, we find that $\iota_{\rm max}$ has a very sharp transition from $\iota_{\rm max} \approx 0$ when $r_0=3M$ to $\iota_{\rm max} \approx \pi/2$ at infinity, with the transition taking place around $r_0=4M$. 
\begin{figure}[t]
\centering
\includegraphics[width=.47 \textwidth]{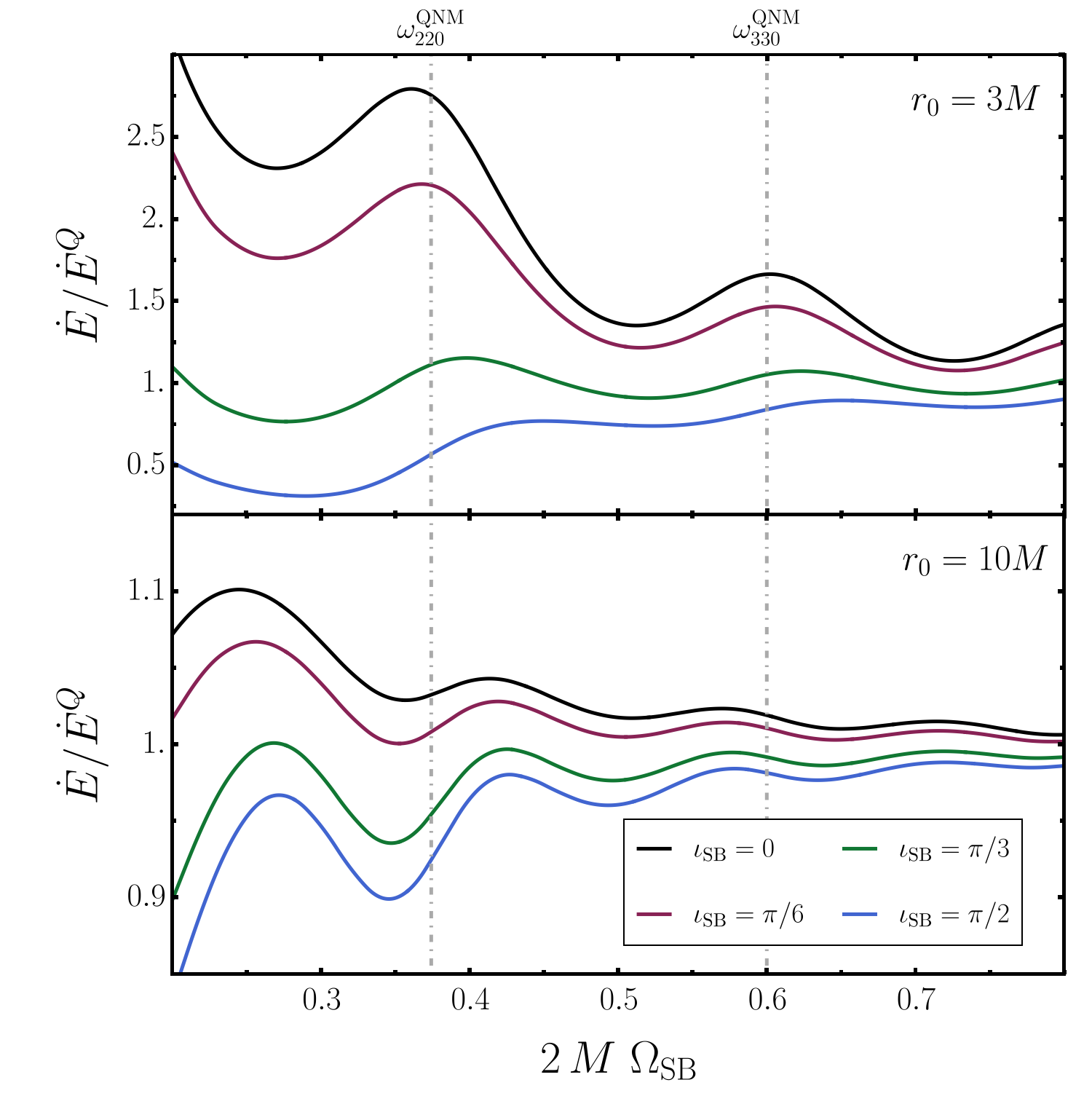}
\caption{\justifying
Total energy flux $\dot E = \dot E^\infty +\dot E^H$, normalized to $\dot E ^Q$, for a SB at a distance $r_0$ from the SMBH and with intrinsic spin at various angles $\iota_{\rm SB}$ relative to the $\hat z$-axis (geometric setup in Fig.~\ref{fig:InclinedGeometry}).
{\bf Top}: SB at the light ring $r_0=3M$; {\bf Bottom}: SB at $r_0=10 M$.
In both plots, the flux is larger when the inclination is zero, but approaches the quadrupole formula in the geometric-optics limit for all values of $\iota_{\rm SB}$.
When the binary is at the light ring, the resonances (peaks) are most prominent when $\iota_{\rm SB}=0$, as in that case the spin is tangent to the light ring.
By contrast, for $r_0=10M$, resonances (peaks) are most prominent when $\iota_{\rm SB}=\pi/2$.
We find that $\iota_{\rm SB}\approx0$ is optimal for $r_0\lesssim4M$, while $\iota_{\rm SB}\approx\pi/2$ maximizes the resonance when $r_0\gtrsim4M$.}
\label{fig:InclinedTotal}
\end{figure}

Let us then revisit the cases of a static CoM SB at $r_0=3M$ and $r_0=10M$, but with the intrinsic spin at an inclination $\iota_{\rm SB}\neq0$. Note that by inclining the binary the reflection symmetry referred to in the last paragraph of Sec.~\ref{sec:b_EMRI} is broken. In Fig.~\ref{fig:InclinedTotal} we show the total flux $\dot E = \dot E^\infty + \dot E ^H$ for varying inclinations $\iota_{\rm SB}$. In the geometric optics limit $\omega \equiv 2 \Omega_{\rm SB} \to \infty$, the total flux approaches the quadrupole formula~\eqref{eq:EQ}, as expected. 

We want to highlight three points in Fig.~\ref{fig:InclinedTotal}. First, when the binary is pointing in the direction of the SMBH the system emits less energy. This is likely due to complicated wave optics phenomena, as the difference decreases with increasing frequency. Second, the excitation of modes (peaks) is sensitive to the value of $\iota_{\rm SB}$. In particular for $r_0 =3M$ they are more prominent when $\iota_{\rm SB}=0$; this is to be expected, as in that case the spin of the SB is tangent to the light ring. In contrast, for $r_0 =10M$ the resonance is maximal when  $\iota_{\rm SB}=\pi/2$; this has a more complicated interpretation, given in Appendix~\ref{app:Inclined}, and is true for any SB at a distance $r_0 \gtrsim 4M$ from the SMBH. Finally, we note that the frequency that maximizes the resonant excitation of modes depends on the inclination of the SB.   

%
%%%%%%%%%%%%%%%%%%%%%%%%%%%%%%%%%%%%%%%%%%
\subsection{The effect of the supermassive BH spin} \label{sec:kerr}
%%%%%%%%%%%%%%%%%%%%%%%%%%%%%%%%%%%%%%%%%%
%
\begin{figure}[t]
\centering
\includegraphics[width=.47 \textwidth]{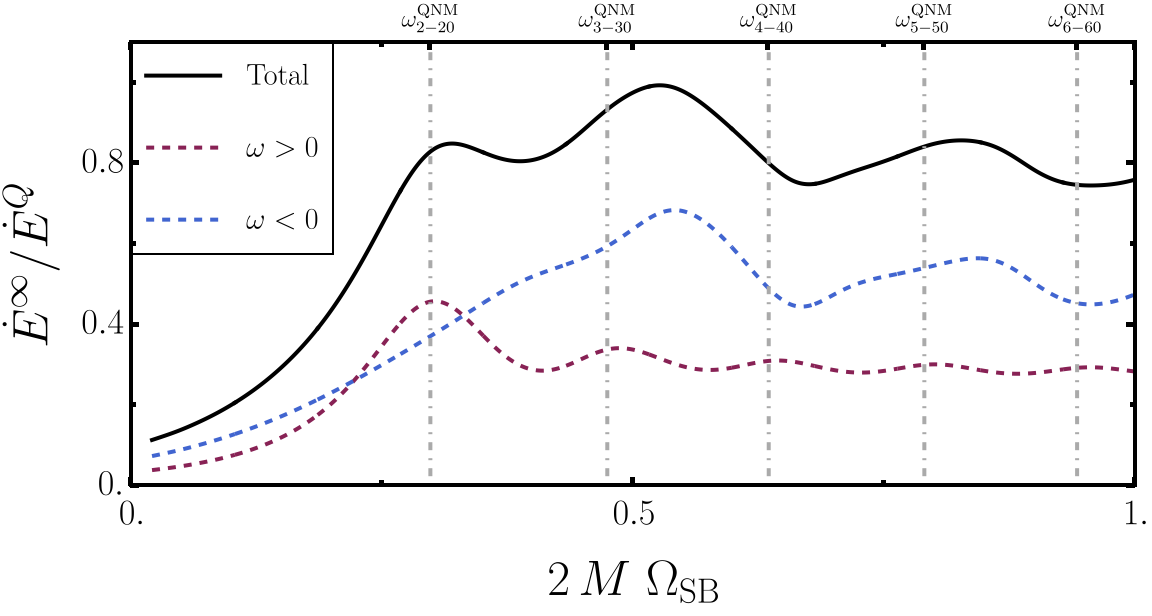}
\caption{\justifying
Energy flux at infinity $\dot E^\infty$, normalized to $\dot E ^Q$, for a static CoM SB at the retrograde light ring in the equatorial plane of a Kerr BH with spin $a = 0.9 M$.
The retrograde light ring is at $r\approx3.9M$, still outside the ergoregion ($r<2 M$).
The intrinsic spin of the SB points along the light ring, similar to SB II in Fig.~\ref{fig:BinaryGeometry}.
This setup breaks the usual symmetry of the our b-EMRI model under reflections about the equatorial plane, which means that modes with $\omega>0$ ($q=+2$) and $\omega<0$ ($q=-2$) are independent of each other (see Eq.~\eqref{eq:frequencies}).
The dashed lines show the energy flux only in $\omega>0$ modes (red) and $\omega<0$ modes (blue); combining the two yields the black curve.
The positive-(negative-)frequency modes capture the radiation emitted (anti-)parallel to the SB intrinsic spin.
Thus, the positive-frequency modes feed the light ring directly, as we saw for Schwarzschild in Fig.~\ref{fig:Sch_LR_total}.
In contrast, the negative-frequency modes are emitted mostly in a prograde direction, and hence do not feed the local light ring; this leads to an unpredictable resonance profile, more similar to Fig.~\ref{fig:Sch_r=10} but with the added difficulty that the Kerr has a far denser resonance spectrum.
The overall result is that the resonances are harder to assign to individual modes when the SMBH is rotating.}
\label{fig:Kerr}
\end{figure}

Having thoroughly investigated the phenomenon of excitation of QNMs of non-spinning BHs, we now briefly visit the generic case of spinning BHs, described by the Kerr metric. Spinning BHs have longer lived QNMs. This means that the resonance should be sharper and better described by the physics of undamped oscillators (Lorentzian curves, etc). On the other hand, their QNM frequencies now depend on the azimuthal number $m$. This makes their spectrum far denser in the frequency plane, possibly making individual resonances harder to identify.

To test these hypotheses, we place a stationary CoM SB (at rest with respect to the orbits of $\partial_t$) in the retrograde light ring in the equatorial plane of a Kerr BH with spin $a = 0.9 M$. The retrograde light ring is at $r\approx3.9M$, still outside the ergoregion ($r<2 M$). We also rotate the SB so that its intrinsic spin points along the light ring, analogously to SB II in Fig.~\ref{fig:BinaryGeometry}, in the retrograde direction (that is, in the same direction as the null geodesics).

In Fig.~\ref{fig:Kerr} we look at the energy flux at infinity for this source, as a function of the SB frequency. This setup breaks the reflection symmetry about the $z=0$ plane. Thus, the amplitudes don't have the symmetry mentioned in the last paragraph of Sec.~\ref{sec:b_EMRI}, and modes with positive and negative frequency display different behaviors. In particular, the former are mostly emitted along the SB intrinsic spin (directly feeding the retrograde light ring as in Sec.~\ref{sec:LR}), while the latter are emitted opposite to it.
For this reason, the modes with $\omega>0$ (red dashed) show clear resonance with the retrograde light ring modes $m=-\ell$, contrary to the modes with $\omega<0$ (blue dashed). The overall flux to infinity (black) is dominated by the negative frequency modes, as the positive frequency modes are more absorbed by the BH; thus, the total flux to infinity does not show a clear resonance with retrograde (or any other) modes.\footnote{We don't show the fluxes on the horizon as these don't give any new information about the system. Qualitatively, their behavior can be inferred from Fig.~\ref{fig:Kerr} and the bottom panel of Fig.~\ref{fig:Sch_LR_total}.}
%
%%%%%%%%%%%%%%%%%%%%%%%%%%%%%%%%%%%%%%%%%%
\section{Discussion} \label{sec:Discussion}
%%%%%%%%%%%%%%%%%%%%%%%%%%%%%%%%%%%%%%%%%%
%
General Relativity exhibits a remarkably rich mathematical structure and a wide variety of solutions.
In particular, BH spacetimes possess an event horizon, photon spheres, and characteristic modes of oscillation.
The primary goal of this work was to understand the excitation of these modes in a realistic astrophysical setting equipped with a controllable ``knob'' that allows one to sweep through frequency space.
Such a system is naturally provided by a neighboring stellar-mass binary whose intrinsic orbital frequency evolves over time and may transiently resonate with the SMBH.
In this sense, the stellar-mass binary can act as a gravitational tuning fork, exciting the oscillation modes of a nearby SMBH.
The main conclusion of our study is that stellar-mass binaries can indeed excite the QNMs of a SMBH, as evidenced by variations in the energy fluxes measured both at infinity and on the horizon.

We focused on the idealized configuration in which the center of mass (CoM) of the stellar-mass binary is held fixed, primarily because this renders the GW signal monochromatic and greatly simplifies the analysis.
While this setup may appear only marginally realistic, it is worth emphasizing that ringdown timescales are governed by light-ring properties and are therefore much shorter than typical orbital timescales.
As a result, the static-CoM approximation is expected to capture the essential features of the resonant response.

We began by considering a non-spinning SMBH and a secondary binary (SB) whose intrinsic angular momentum is oriented perpendicular to the line connecting the binary to the SMBH.
In this configuration, the resonant excitation of BH modes is particularly clear.
We find that the resonance occurs at a frequency close to---but not exactly equal to---the real part of the corresponding QNM frequency.
Moreover, as the distance of the SB from the SMBH increases (while keeping the orientation of its intrinsic spin fixed), the frequency that maximizes the resonant response drifts further away from the QNM frequency, as illustrated in Fig.~\ref{fig:Sch_multi_total}.
Both effects are key results of this work and reflect generic features of dissipative systems, as exemplified both by BHs and by the toy string model discussed in Appendix~\ref{sec:Toy}.
As expected, the resonance is strongest when the SB is located near the light ring, where QNMs are effectively trapped.

Additionally, we argued that, within the quadrupole approximation adopted for the SB, most of the radiation is emitted along the direction of its intrinsic spin.
Consequently, when this direction is aligned so as to maximize the radiation incident on the light-ring region of the SMBH, the excitation of QNMs can be significantly enhanced, as shown in Fig.~\ref{fig:InclinedTotal}.

We also attempted to quantify the mode excitation in terms of the energy flux carried by individual harmonic modes.
However, individual harmonics are not frame-independent: different modes are related by frame rotations and translations, and changing the orientation of the SB relative to the SMBH modifies the relative excitation of different azimuthal modes.
For these reasons, we found no robust signatures of QNM excitation at the level of individual harmonics.
Indeed, as discussed in Sec.~\ref{sec:Flat}, even in flat spacetime, the excitation of individual multipoles can be highly nontrivial when the source is not located at the origin of the coordinate system.

The resonant response of spinning SMBHs is substantially more difficult to disentangle due to the richer structure of their QNM spectrum.
In the Schwarzschild limit, modes with different azimuthal numbers $m$ are degenerate for a given angular index $\ell$, whereas for Kerr BHs this degeneracy is lifted.
The resulting denser spectrum makes individual resonances harder to isolate, as illustrated in Fig.~\ref{fig:Kerr}.
On the other hand, QNMs of rapidly spinning SMBHs are less strongly damped, which may lead to sharper resonances if individual modes can be identified.
From the perspective of the stellar-mass binary, a spinning SMBH resembles an ellipsoidal piano, with different orientations exciting different notes (cf. Sec.~\ref{sec:Piano}).
We leave a detailed exploration of these effects for future work, so as not to obscure the main focus of this article, which we believe establishes a solid foundation for the study of resonant QNM excitation by b-EMRIs.

Finally, several important extensions of this work remain to be explored.
These include the excitation of overtones, the possible role of nonlinear effects, resonant phenomena when the SB is not held static, and the detectability of such resonances.
Regarding this last point, changes in the energy flux modify the evolution rate of the stellar-mass binary and may be observable if the system lies within the sensitivity band of a low-frequency detector such as LISA.
However, modeling such effects would require a self-consistent treatment of backreaction, for instance through a flux-balance law.
An alternative possibility is the direct detection of an excess in the gravitational-wave amplitude using higher-frequency ground-based detectors.

%%%%%%%%%%%%%%%%%%%%%%%%%%%%%%%%%%%%%%%%%%
\section{Acknowledgments}
%%%%%%%%%%%%%%%%%%%%%%%%%%%%%%%%%%%%%%%%%% 
%
We thank Ana Carvalho for producing Figs.~1 and 2.
The Center of Gravity is a Center of Excellence funded by the Danish National Research Foundation under grant no.\ DNRF184.
We acknowledge support by VILLUM Foundation (grant no.\ VIL37766).
VC and J.S.S. acknowledge financial support provided under the European Union’s H2020 ERC Advanced Grant “Black holes: gravitational engines of discovery” grant agreement no.\ Gravitas–101052587.
V. C. and J. S. S. also thank the Fundação para a Ciência e Tecnologia (FCT), Portugal, for the financial support to the Center for Astrophysics and Gravitation (CENTRA/IST/ULisboa) through Grant No. UID/PRR/00099/2025 and Grant No. UID/00099/2025.
AL was supported in part by NSF grant AST-2307888 and the NSF CAREER award PHY-2340457.
JN was partially funded by Fundação para a Ciência e Tecnologia (FCT), Portugal, through grant no.\ UID/4459/2025, and by the European Union’s Horizon 2020
research and innovation programme H2020-MSCA-2022-SE
through project EinsteinWaves, grant agreement no.\ 101131233.
MvdM acknowledges financial support provided under the European Union’s Horizon ERC Synergy Grant “Making Sense of the Unexpected in the Gravitational-Wave Sky” grant agreement no.\ GWSky–101167314.
Views and opinions expressed are however those of the authors only and do not necessarily reflect those of the European Union or the European Research Council. Neither the European Union nor the granting authority can be held responsible for them.
This project has received funding from the European Union's Horizon 2020 research and innovation programme under the Marie Sk{\l}odowska-Curie grant agreement No 101007855 and No 101131233, 
as well as from Funda\c{c}\~{a}o para a Ci\^{e}ncia e a Tecnologia under the project 2024.04456.CERN.
This work is supported by Simons Foundation International \cite{sfi} and the Simons Foundation \cite{sf} through Simons Foundation grants SFI-MPS-BH-00012593-09 and SFI-MPS-BH-00012593-11.

\newpage
\appendix

%%%%%%%%%%%%%%%%%%%%%%%%%%%%%%%%%%%%%%%%%%%%%%%%%%%%%%%
\section{A toy model for QNM excitation} \label{sec:Toy}
%%%%%%%%%%%%%%%%%%%%%%%%%%%%%%%%%%%%%%%%%%%%%%%%%%%%%%%
%
%%%%%%%%%%%%%%%%%%%%%%
\subsection{The model}
%%%%%%%%%%%%%%%%%%%%%%
%
A string attached at both ends is a well-known example of a conservative system.
If one end is allowed to dissipate, then its modes become quasinormal, even if the equation describing the system is still the wave equation.

Consider then the boundary value problem for the wave equation given by
\begin{eqnarray}
&&\frac{d^2\psi}{dx^2}+\omega^2\psi=0\,.\\
&&\psi(0)=0\,,\quad \psi'(L)=i\omega\delta \psi(L)\,,
\end{eqnarray}
where primes are derivatives with respect to argument and $\delta$ is a constant.
This represents a string fixed at $x=0$ and dissipating at $x=L$.\footnote{More precisely, this is a Fourier transform in time of the system \[\partial^2_t \psi = \partial^2_x \psi \, , \quad \psi(t,0)=0 \, , \quad \partial_x \psi(t,L) = -\delta \partial_t \psi(t,L) \, ,\] whose energy decreases, since \[\frac{d}{dt} \int_0^L \frac12 \left( (\partial_t \psi)^2 + (\partial_x \psi)^2 \right) dx = - \delta (\partial_t \psi(t,L))^2 \, .\]}
The solution is of the form $\psi=Ae^{-i\omega x}+Be^{i\omega x}$.
Imposing the boundary conditions, we find
\begin{equation}
-1=\frac{1-\delta}{1+\delta}e^{2i\omega L}\,,
\end{equation}
or
\begin{equation}
L\omega=L\omega^{\rm QNM} _ n=\pi(n+1/2)-\frac{i}{2}\log\frac{1+\delta}{1-\delta}\,.
\end{equation}
We see that dissipation introduces an imaginary part to the frequencies, making the modes quasinormal.
Indeed, in the small-$\delta$ limit, $L\omega^{\rm QNM} _ n=\pi(n+1/2)-i\delta$.

Now let us excite this string with a monochromatic pulse of frequency $\omega_R$ by a delta-like source at $x=x_0$.
One finds that the problem to be solved is now\footnote{To obtain a real solution, we should also add a term of the form $\delta(x-x_0)\delta(\omega+\omega_R)$.
Here, we will instead consider a complex solution and then take the real part.}
\begin{eqnarray}
&&\frac{d^2\psi}{dx^2}+\omega^2\psi=\delta(x-x_0)\delta(\omega-\omega_R)\,,\\
&&\psi(0)=0\,,\quad \psi'(L)=i\omega\delta \psi(L)\,,
\end{eqnarray}
whose solution is
\begin{equation}
\psi = \Big(\psi_< \Theta(x_0-x) + \psi_> \Theta(x-x_0)\Big) \delta(\omega - \omega_{\rm R})\,,
\end{equation} 
with $\Theta$ is the Heaviside step function and 
\begin{eqnarray}
\psi_< (x, \omega)&=&\frac{\sin \left(\omega x \right) \left(\cos (\omega  (L-x_0))-i \delta  \sin (\omega  (L-x_0))\right)}{-\omega \left(
  \cos (L \omega )+i \delta  \sin (L \omega ) \right)} \, , \nonumber \\
\psi_>(x,\omega)&=&\frac{\sin (\omega x_0 )\left(\cos (\omega  (L-x))-i \delta  \sin (\omega  (L-x))\right)}{-\omega \left(
  \cos (L \omega )+i \delta  \sin (L \omega ) \right)}  \, . \nonumber
\end{eqnarray}
%
%%%%%%%%%%%%%%%%%%%%%%%%%%%%%%%%%%%%%%%%
\subsection{Resonances in the toy model}
%%%%%%%%%%%%%%%%%%%%%%%%%%%%%%%%%%%%%%%
%
The response of the system to a driving force depends on the driving frequency $\omega_R$.
In the absence of damping, $\delta=0$, the amplitude of the oscillations has a pole when $\omega_R = \omega^{\rm QNM} _ n$ (recall that in this case the modes are normal so $\omega^{\rm QNM} _ n$ is real).
By contrast, if $\delta \neq0$, then this is no longer the case, and now the resonance is maximal for $\omega_R = \omega_{\rm max}(\delta,x_0)$. 

To illustrate this point, we need a tool to diagnose the strength of the resonance.
By analogy with the observables we use in GW physics, we consider the energy flux out of the system at $x=L$, which is given by\footnote{The time-average of $(\operatorname{Re} (-i\omega_R \psi_> e^{-i\omega_R t}))^2$, or equivalently of $\frac14 (-i\omega_R \psi_> e^{-i\omega_R t} + i \omega_R \psi_>^* e^{i\omega_R t} )^2$, is precisely $\omega_R^2\frac{|\psi_>|^2}2$.}
\begin{equation}
\dot{E} = \delta \omega_R ^2  \frac{|\psi_>(L, \omega_R)|^2}2 \, .
\end{equation}
For the model above,
\begin{equation}
 \dot{E}=\frac{1}{2}\frac{\delta \sin^2{(\omega_R x_0)}}{\cos^2{(\omega_R L)}+\delta^2\sin^2{(\omega_R L)}}\,.
\end{equation}
It is clear from this expression that the flux has sharp maxima at $\omega_R=\omega_n^{\rm QNM}$ when the damping $\delta$ is small.
However, at any finite $\delta$, the local maxima in the flux depend both on $x_0$ and $\delta$, and are not, in general, at $\omega_R={\rm Re}[\omega^{\rm QNM}_n]$.
In the limit $\delta\to 1$, the flux peaks at a frequency $\omega_R=k\pi/(2x_0)$, which depends on $x_0$ alone.

\begin{figure}[t]
\centering
\includegraphics[width=.45 \textwidth]{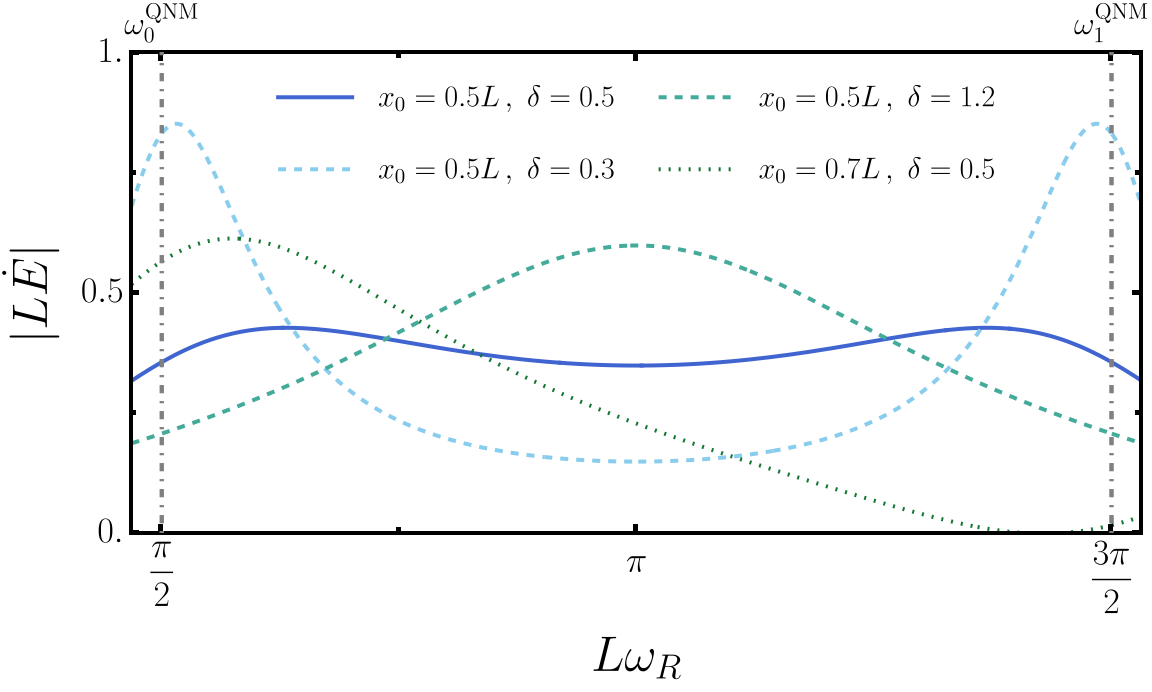}
\caption{\justifying
Energy flux at $x=L$ as a function of the driving frequency $\omega_R$, for the toy model of a string with length $L$ and dissipation parameter $\delta$, being forced at position $x_0$.
Different curves correspond to different values of $\delta$ and $x_0$.
Vertical lines indicate the real parts of the first two QNM frequencies.
Varying $\delta$ and/or $x_0$ shifts the frequency $\omega_{R}$ that maximizes the response of the system, which is generally \emph{not} the frequency of any QNM.}
\label{fig:Toy}
\end{figure}

In the small-damping regime, the flux peaks at
\begin{eqnarray}
L\omega_R&=&\pi/2+n\pi+Z\delta^2\,, \label{eq:frequencies_toy}\\
Z&=&\frac{x_0}{L}\cot\left(\frac{(2n+1)\pi x_0}{2L}\right)
\end{eqnarray}
It is readily seen that when the source is close to the boundary, $x_0\sim L$, and the overtone number is low, then $Z>0$ and the frequency of maximum flux is {\it larger} than the QNM frequency of the system.
To further illustrate this point, we show in Fig.~\ref{fig:Toy} the energy flux as a function of $\omega_R$ for different values of $\delta $ and $x_0$. 

Any system with energy dissipation has complex characteristic frequencies, called quasinormal modes $\omega^{\rm QNM}$. By analyzing the simple dissipative toy model of a string with dissipation parameter $\delta$, being forced at $x=x_0$ with frequency $\omega_R$, we conclude: \emph{the response of these systems is maximized for $\omega_R = \omega_{\rm max}(\delta,x_0)$, where we generally have $\omega_{\rm max}\neq {\rm Re}[\omega^{\rm QNM}]$ for all $\delta \neq 0$.} 
%
%%%%%%%%%%%%%%%%%%%%%%%%%%%%%%%%%%
\section{Formalism for translated binary in flat space} \label{app:FlatFormalism}
%%%%%%%%%%%%%%%%%%%%%%%%%%%%%%%%%%
%
Here, we outline how the results in Sec.~\ref{sec:Flat} were obtained.
The idea is to solve the Teukolsky equation for $s=-2$ and set $M=a=0$.
We decompose the field into Fourier-harmonic components, just as in Eq.~\eqref{eq:psi4_inf}, which yields
\begin{equation}
    \psi_4 \sim \sum_{\substack{\ell, m ,q}} {Z}^N _{\ell m q} \frac{e^{-i\omega_{00q} u}}{r} \, _{-2}Y _{ \ell m} (\theta) e^{i m \phi} \, . \label{eq:psi4_inf_flat}
\end{equation}
Since the CoM is now static, the only frequencies present are $\omega_{00\pm2}$ in Eq.~\eqref{eq:frequencies}, as expected within the quadrupole approximation.
The spherical symmetry of flat spacetime implies that the angular eigenfunctions are now spin-weighted \emph{spherical} harmonics. The amplitudes are determined by
\begin{equation}
    Z_{\ell m q}^N = \sum_ 
    {i= 0} ^{4}\ \sum_ 
    {j = 0} ^{4-i} \mathcal{B}^{(i,j)}_{\ell m q} \frac{d^i }{dr^i} R^{N}(r_0)\frac{d^j}{d\theta^j}  \bar{Y}(\pi/2) \, , \label{eq:amplitudes_flat}
\end{equation} 
where, as in the main text, $\mathcal{B}$ depends only on the outer and inner orbital parameters of the SB, $\bar Y \equiv {_{-2}\bar{Y}_{\ell m }}$, and $R^{N}\equiv R^{N}_{\ell m q} \equiv R^{N}_{\ell m \omega_{00q}}$ is a solution to the homogeneous radial equation,
\begin{align}
    & \left(\frac{d^2 }{dr^2}-\frac{2 }{r}\frac{d}{dr}+V(r)\right) R^N_{\ell m \omega} = 0 \, \label{eq:Teukolsky_flat} \, , \\ 
  & V(r) = \omega^2 - \frac{4 i \omega}{r} -\frac{\ell(\ell+1)-2}{r^2} \, .
\end{align}
We wish to find solutions to this equation that are regular at the origin $r=0$.
To do so, we first recast the problem in a simpler form using a Chandrasekhar transformation~\cite{Chandrasekhar:1975nkd}.
The latter relates solutions to the Teukolsky equation for $a=0$ (a more generic class than we are interested in here) to solutions of the Regge-Wheeler equation~\cite{Regge:1957td}, $X_{\ell m \omega}$, which for $M=0$ is just the wave equation: 
\begin{align}
    & \left(\frac{d^2}{d r^2} + \omega^2 -\frac{\ell(\ell+1)}{r^2}\right) \, X_{\ell m \omega} =0 \label{eq:RW_flat}\, .
\end{align}
Then, to obtain the solution of Eq.~\eqref{eq:Teukolsky_flat}, we apply the Chandrasekhar transformation:
\begin{equation}
    R^N _{\ell m \omega} = r^2 \left( \frac{d}{dr} + i \omega\right)^2 (r X^H _{\ell m \omega}) \, .
\end{equation}
We want solutions to Eq.~\eqref{eq:RW_flat} which are regular at the origin, so we must have 
\begin{equation}
    X^H_{\ell m \omega} (r) = \omega r \,  j_\ell (\omega r) = \eta \, j_\ell (\eta) \, ,
\end{equation}
where we have defined $\eta\coloneq \omega r $, and $j_\ell (\eta)$ are spherical Bessel functions of the first kind.
Following this recipe, it is straightforward to obtain the amplitudes using Eq.~\eqref{eq:amplitudes_flat}, though we emphasize that $\mathcal{B}$ is still very complicated, since the source term is not simple.
%
%%%%%%%%%%%%%%%%%%%%%%%%%%%%%%%%%%
\section{Consistency checks for inclined binaries} \label{app:Inclined}
%%%%%%%%%%%%%%%%%%%%%%%%%%%%%%%%%%
%
\begin{figure}[t]
\centering
\includegraphics[width=.3 \textwidth]{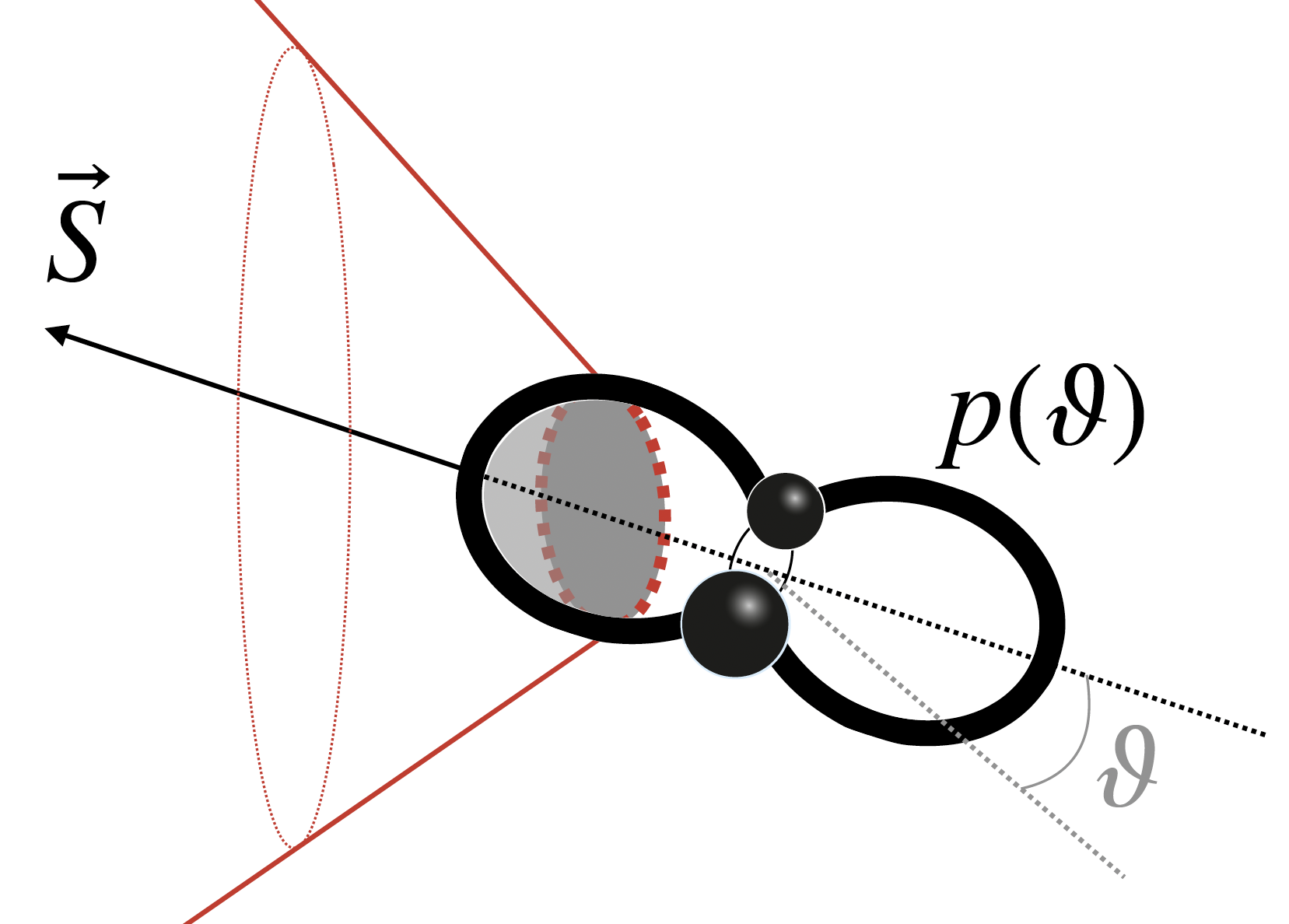}
\caption{\justifying
Geometric setup in the local frame of an inclined SB at a distance $r_0$ from the SMBH.
The critical null cone is shown in red (compare with Fig.~\ref{fig:InclinedGeometry}) and the intrinsic spin of the SB as a black arrow.
The thick black line represents the angular distribution of radiated energy by the SB in its local rest frame, in the geometric-optics limit.
The shape of this angular distribution is given in Eq.~\eqref{eq:PM} as a function of $\vartheta$, the angle between a direction and the spin.
The radiation emitted inside of the null cone is absorbed by the BH (gray shaded area) while the rest escapes to infinity.
Radiation emitted along the red dashed circumference gets trapped at the light ring.}
\label{fig:LocalGeometry}
\end{figure}

In this appendix, we show some consistency checks for the simulations of inclined SBs presented in Sec.~\ref{sec:Inclined}.
The analytical results we compare against are all derived within the geometric-optics approximation, which for the static CoM SB systems corresponds to $2 M \Omega_{\rm SB}\gg1$. 

We are interested in systems in which the spin of the SB lies at an angle $\iota_{\rm SB}$ relative to the $\hat{z}$-axis (see Fig.~\ref{fig:InclinedGeometry}).
We work within the quadrupole approximation for the radiation emitted by the inner motion of the SB.
Thus, in a local frame attached to the SB, its differential energy flux is given by the Peters-Matthews formula~\cite{Peters:1963ux}
\begin{equation}
    \frac{d E}{d\Omega} =p(\vartheta) \propto (1 + 6 \cos^2 \vartheta + \cos^4 \vartheta) \, , \label{eq:PM}
\end{equation}
where $\vartheta$ is the angle between any direction and the intrinsic spin of the SB (see  Fig.~\ref{fig:LocalGeometry}).

\begin{figure}[t]
\centering
\includegraphics[width=.47 \textwidth]{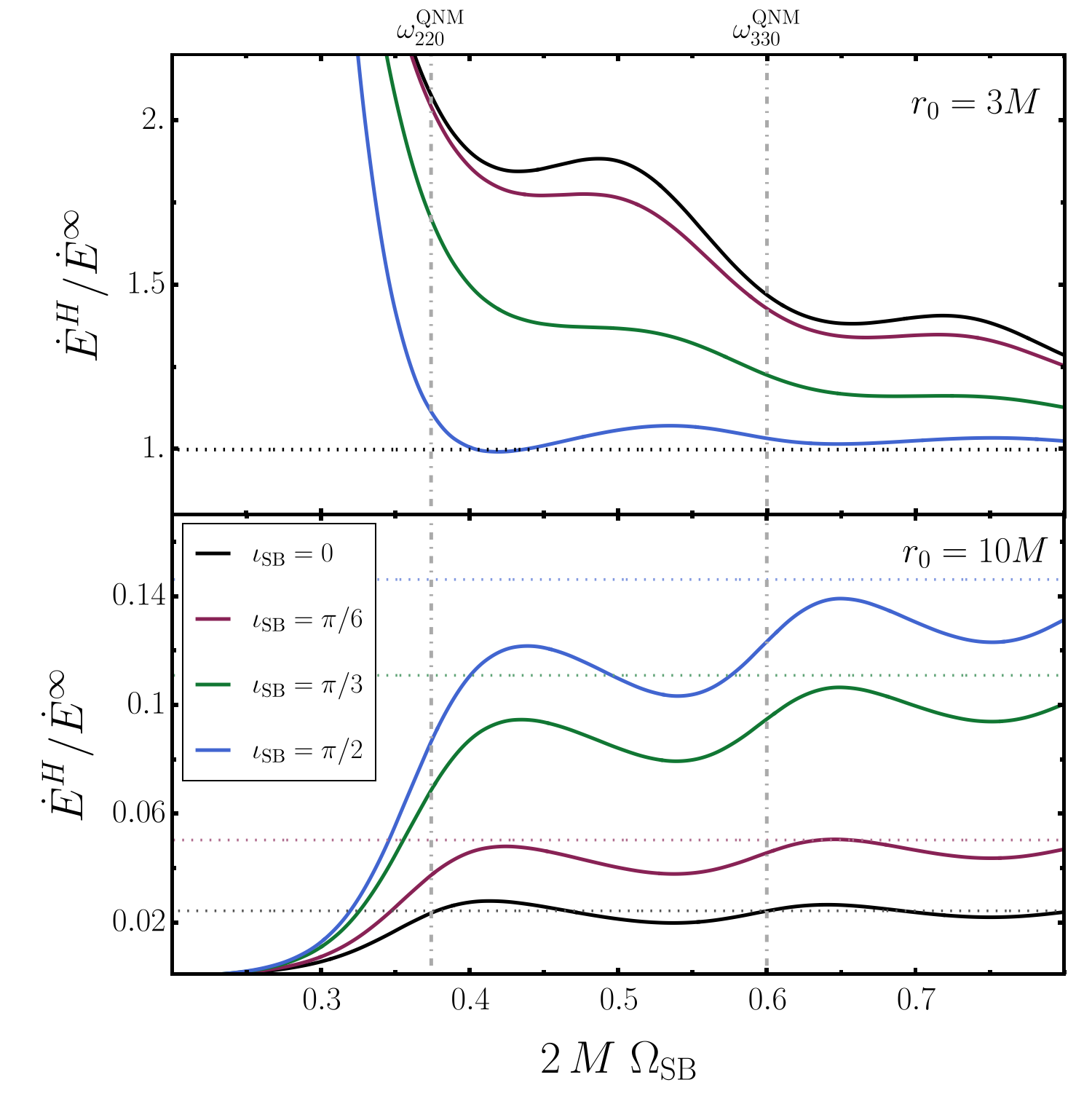}
\caption{\justifying
Ratio of energy flux on the horizon, $\dot E^H$, to the flux at infinity, $\dot E^\infty$, in the same setup as in Fig.~\ref{fig:InclinedTotal}.
{\bf Top}: Binary at the light ring $r_0=3M$.
{\bf Bottom}: Binary at $r_0=10M$.
Different colors indicate different inclinations of the SB intrinsic spin $\iota_{\rm SB}$ as illustrated in Fig.~\ref{fig:InclinedGeometry}.
Dotted horizontal lines show the analytical predictions obtained in the geometric-optics limit $2M\Omega_{\rm SB}\gg1$.
The numerical and analytical results appear to agree in this regime.}
\label{fig:InclinedRatio}
\end{figure}
We first want to develop a model for how much radiation is absorbed by the BH compared to the energy that escapes to infinity. In the geometric optics limit, radiation emitted to the inside of the critical null cone will be absorbed by the BH. This can be computed by integrating Eq.~\eqref{eq:PM} over the gray shaded area in Fig.~\ref{fig:LocalGeometry}. The rest of the energy will escape to infinity. In Fig.~\ref{fig:InclinedRatio} we investigate the value of the ratio $\dot E^H /\dot E^\infty $ for the same b-EMRI setups that we studied in Sec.~\ref{sec:Inclined}. We compare these with the result obtained by calculating the integral mentioned above (given by the dotted horizontal lines). We see that the results tend to agree at large frequencies, which is precisely the regime where the analytical estimate is derived. 

Finally, we want to comment on the inclination  $\iota_{\rm SB}=\iota_{\rm max}(r_0)$ which maximizes the resonant behavior. For each distance $r_0$ from the SB to the SMBH, we expect $\iota_{\rm max}$ to be such that the average of $p(\vartheta)$ over the intersection of the null cone with a sphere centered at the SB (red dashed circle in Fig.~\ref{fig:LocalGeometry}) is maximal. We can calculate this by integrating Eq.~\eqref{eq:PM} over this intersection for various inclinations. We find that $\iota_{\rm max} \approx 0$ for $r_0<3.7M$ and $\iota_{\rm max} \approx \pi/2$ for $r_0>4.1M$, with a very sharp transition around $r_0=4M$.

%%%%%%%%%%%%%%%%%%%%%%%%%%
%

\end{document}